\documentclass[
reprint,
 aps,
 superscriptaddress,
bibnotes,
pra
]{revtex4-2}
\usepackage{physics}
\usepackage{dsfont}
\usepackage{color,newfloat}
\usepackage{graphicx}
\usepackage{dcolumn}
\usepackage{bm}
\usepackage{amsmath,amssymb,amsthm}
\usepackage{mathtools}
\usepackage{multirow}
\usepackage{hhline}
\usepackage{comment}
\usepackage[dvipsnames]{xcolor}
\usepackage{hyperref}
\hypersetup{
    colorlinks=true,
    linkcolor=Maroon,
    citecolor=Maroon,
    urlcolor=Maroon
}
\usepackage{gensymb} 

\usepackage[caption = false]{subfig}
\usepackage{lineno}
\usepackage{siunitx}
\usepackage[mathscr]{euscript}
 \let\mathscr\relax

\usepackage{makecell}

\begin{document}


\title{
Programmable Nonlinear Quantum Photonic Circuits}

\author{Kasper H. Nielsen}
\altaffiliation{Equally contributing authors.}
\affiliation{Center for Hybrid Quantum Networks (Hy-Q), Niels Bohr Institute, University of Copenhagen, Blegdamsvej 17, Copenhagen 2100, Denmark}

\author{Ying Wang}
\altaffiliation{Equally contributing authors.}
\affiliation{Center for Hybrid Quantum Networks (Hy-Q), Niels Bohr Institute, University of Copenhagen, Blegdamsvej 17, Copenhagen 2100, Denmark}

\author{Edward Deacon}
\affiliation{Quantum Engineering and Technology Laboratories, School of Physics and Department of Electrical and Electronic Engineering, University of Bristol, Bristol, UK}

\author{Patrik I. Sund}
\affiliation{Center for Hybrid Quantum Networks (Hy-Q), Niels Bohr Institute, University of Copenhagen, Blegdamsvej 17, Copenhagen 2100, Denmark}

\author{Zhe Liu}
\affiliation{Center for Hybrid Quantum Networks (Hy-Q), Niels Bohr Institute, University of Copenhagen, Blegdamsvej 17, Copenhagen 2100, Denmark}

\author{Sven Scholz}
\affiliation{Lehrstuhl f{\"u}r Angewandte Festk{\"o}rperphysik, Ruhr-Universit{\"a}t Bochum, Universit{\"a}tsstrasse 150, D-44780 Bochum, Germany}

\author{Andreas D. Wieck}
\affiliation{Lehrstuhl f{\"u}r Angewandte Festk{\"o}rperphysik, Ruhr-Universit{\"a}t Bochum, Universit{\"a}tsstrasse 150, D-44780 Bochum, Germany}

\author{Arne Ludwig}
\affiliation{Lehrstuhl f{\"u}r Angewandte Festk{\"o}rperphysik, Ruhr-Universit{\"a}t Bochum, Universit{\"a}tsstrasse 150, D-44780 Bochum, Germany}

\author{Leonardo Midolo}
\affiliation{Center for Hybrid Quantum Networks (Hy-Q), Niels Bohr Institute, University of Copenhagen, Blegdamsvej 17, Copenhagen 2100, Denmark}

\author{Anders S. Sørensen}
\affiliation{Center for Hybrid Quantum Networks (Hy-Q), Niels Bohr Institute, University of Copenhagen, Blegdamsvej 17, Copenhagen 2100, Denmark}

\author{Stefano Paesani}
\email{stefano.paesani@nbi.ku.dk}
\affiliation{Center for Hybrid Quantum Networks (Hy-Q), Niels Bohr Institute, University of Copenhagen, Blegdamsvej 17, Copenhagen 2100, Denmark}
\affiliation{NNF Quantum Computing Programme, Niels Bohr Institute, University of Copenhagen, Blegdamsvej 17, 2100 Copenhagen, Denmark.}

\author{Peter Lodahl}
\email{lodahl@nbi.ku.dk}
\affiliation{Center for Hybrid Quantum Networks (Hy-Q), Niels Bohr Institute, University of Copenhagen, Blegdamsvej 17, Copenhagen 2100, Denmark}

\begin{abstract}
The lack of interactions between single photons prohibits direct nonlinear operations in quantum optical circuits, representing a central obstacle in photonic quantum technologies.
Here, we demonstrate multi-mode nonlinear photonic circuits where both linear and direct nonlinear operations can be programmed with high precision at the single-photon level.
Deterministic nonlinear interaction is realized with a tunable quantum dot embedded in a nanophotonic waveguide mediating interactions between individual photons within a temporal linear optical interferometer.
We demonstrate the capability to reprogram the nonlinear photonic circuits and implement protocols where strong nonlinearities are required, in particular for quantum simulation of anharmonic molecular dynamics, thereby showcasing the new key functionalities enabled by our technology.
\end{abstract}

\date{\today}

\maketitle


\noindent Photons are carriers of quantum information with unique features for quantum technology: the insensitivity to stochastic noise, the high-speed operation and ability to realize long-distance transmission, and the manufacturability of photonic circuits used for quantum-information processing~\cite{wang2020}. 
Progress in photonic quantum technologies has brought large-scale linear optical quantum photonic circuits consisting of thousands of optical components that process tens of photons~\cite{bao2023}, to realize photonic devices operating in computational regimes that challenge the capabilities of conventional computers for specific quantum computing tasks~\cite{madsen2022,zhong2020}.
However, the lack of direct photon-photon interactions prohibits the nonlinear operations that enable deterministic multi-photon entangling gates~\cite{lindner2009proposal,chang2014quantum} and near-term photonic quantum simulators~\cite{aspuru2012photonic,maring2024versatile}, and currently represents a significant challenge for developing quantum photonic devices at the scale required for practical applications. 
One approach for implementing photon nonlinearities with linear optical circuits relies on measurement-induced operations to mediate photon-photon interactions~\cite{knill2001}.
This approach exploits the nonlinear nature of quantum measurements to reproduce the desired entangling operations~\cite{scheel2003}.
However, because measurements are intrinsically probabilistic in quantum mechanics, such operations necessarily have a finite success probability, dependent on and heralded by the detection of compatible measurement outcomes.  
Multiplexing techniques~\cite{ma2011,mendoza2016active} and quantum error correction~\cite{chiaverini2004realization} methods can be used to turn such probabilistic operations into near-deterministic ones or to tolerate their failures, respectively, but require enormous overheads in terms of hardware components, stringent performance and noise requirements, and classical co-processing~\cite{noiri2022fast,xue2022quantum}.  
As a consequence, while programmable linear optical circuits have been demonstrated in large-scale quantum experiments, near-deterministic nonlinear photonic circuits have so far been elusive for all-optical approaches.
Light-matter interactions in quantum emitters represent an alternative approach to implementing optical nonlinearities at the single-photon level~\cite{fan2010,kojima2003}.
Because photon-photon operations in such systems are mediated through direct interaction with a quantum emitter, this approach enables nonlinearities to be implemented deterministically~\cite{chang2014quantum}.
Recent experiments have demonstrated strong photon-photon nonlinearities with quantum emitters in atomic~\cite{three-photon,nagib2024} and solid-state systems, including quantum dots (QDs) ~\cite{jeannic2022,tomm2023,liu2023violation} and color centers~\cite{pasini2023}, validating the strong potential of this approach for enabling deterministic nonlinear photonic devices. 
Here, we demonstrate a programmable nonlinear photonic quantum processor based on deterministic optical nonlinearities embedded in a programmable linear optical circuit. 
The tunable nonlinearities are implemented on a GaAs chip through the light-matter interaction between single photons and an InAs QD embedded in a photonic-crystal waveguide nanostructure.
These nonlinear operations are combined with programmable linear optical circuits implemented with a temporal interferometer, allowing the use of a single QD nonlinear operation at different times to realize multi-mode nonlinear operations.
We demonstrate separate and simultaneous programmability of both linear and nonlinear operations, thereby displaying how quantum interference and joint spectral properties of the processed photons can be tuned with the different linear and nonlinear control parameters of the circuit.

\begin{figure*}[t]
    \centering
    \includegraphics[width=1\textwidth]{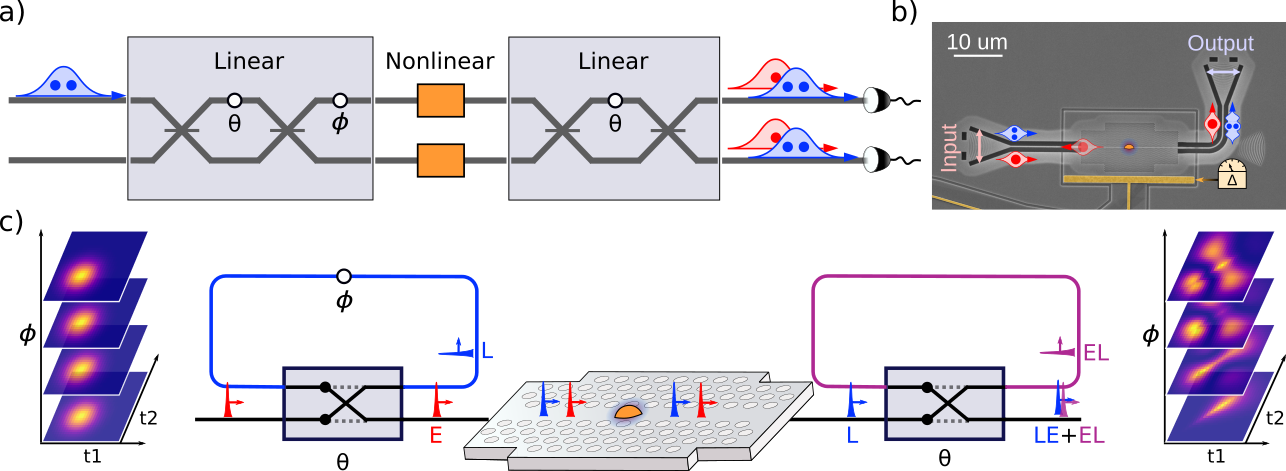}
    \caption{
    \textbf{Schematics of the programmable nonlinear quantum photonic circuit.}
    a) The programmable nonlinear photonic circuit comprises tunable QD nonlinear transformations interleaved between two controllable linear optical circuits implemented via time-bin Mach-Zehnder interferometers that couple a pair of temporal optical modes. 
    The photons are injected into the circuit on the first mode and, subsequently interfere and interact via the linear and nonlinear operations, respectively. The transmitted photons are finally recorded with pseudo-photon-number-resolution using SNSPDs.
    b) SEM image of the GaAs nanophotonic device, which is used to implement the nonlinear interaction by scattering the photons off an InAs QD embedded in a photonic crystal waveguide.
    Grating couplers are used for in- and out-coupling of photons from the device to interconnect it with the linear parts of the circuit.
    The electric field across the QD is controlled with electrodes (highlighted in yellow) that allow tuning of the nonlinear interaction by controlling the  detuning $\Delta$ between the QD resonance and the incoming photons via the DC Stark effect. 
    c) Schematic of the linear optical setup. 
    The pair of optical modes are encoded in the temporal degree of freedom of the photons, where the first optical mode corresponds to an early ($E$) time-bin and the second to a late ($L$) time-bin.  
    The linear operations are implemented in a self-stabilizing time-bin interferometer, which couples the $E$ and $L$ modes and inserts a 
    relative linear phase $\phi$ between the two.
    Photons are injected and retrieved from the GaAs device hosted in a 1.6K cryostat, where nonlinearities are implemented by scattering off the QD at the two different times $E$ and $L$.  
    The two temporal modes are recombined in the time-bin interferometer, where the output interference is affected by both linear and nonlinear phases. 
    The nonlinear interaction significantly changes the properties of the photons, as illustrated by the theoretical joint temporal intensity (JTI) shown with (right inset) and without (left inset) nonlinear interactions for different settings of $\phi$.
    }
    \label{fig:setup}
\end{figure*}
By enabling reliable photon-photon nonlinearities and entangling gates in programmable photonic circuitry, the developed programmable nonlinear quantum photonic platform has potentially transformative applications, e.g., for synthesizing advanced photonic quantum states~\cite{javadi2015single,liang2018observation,jeannic2022} or realizing deterministic Bell-state analyzers ~\cite{witthaut2012photon} and quantum optical neural networks~\cite{steinbrecher2019}.
We illustrate this potential by realizing a proof-of-concept quantum simulation of anharmonic molecular vibrational dynamics by exploiting the natural mapping of this chemistry problem to the nonlinear quantum photonics platform. 
Our work extends pioneering previous work based solely on linear optics and probabilistic measurement-induced nonlinearities~\cite{sparrow2018, alexander2024manufacturable}, thereby paving a way to scaling photonic quantum devices.

\begin{figure*}    
\centering 
\includegraphics[width=1\textwidth]{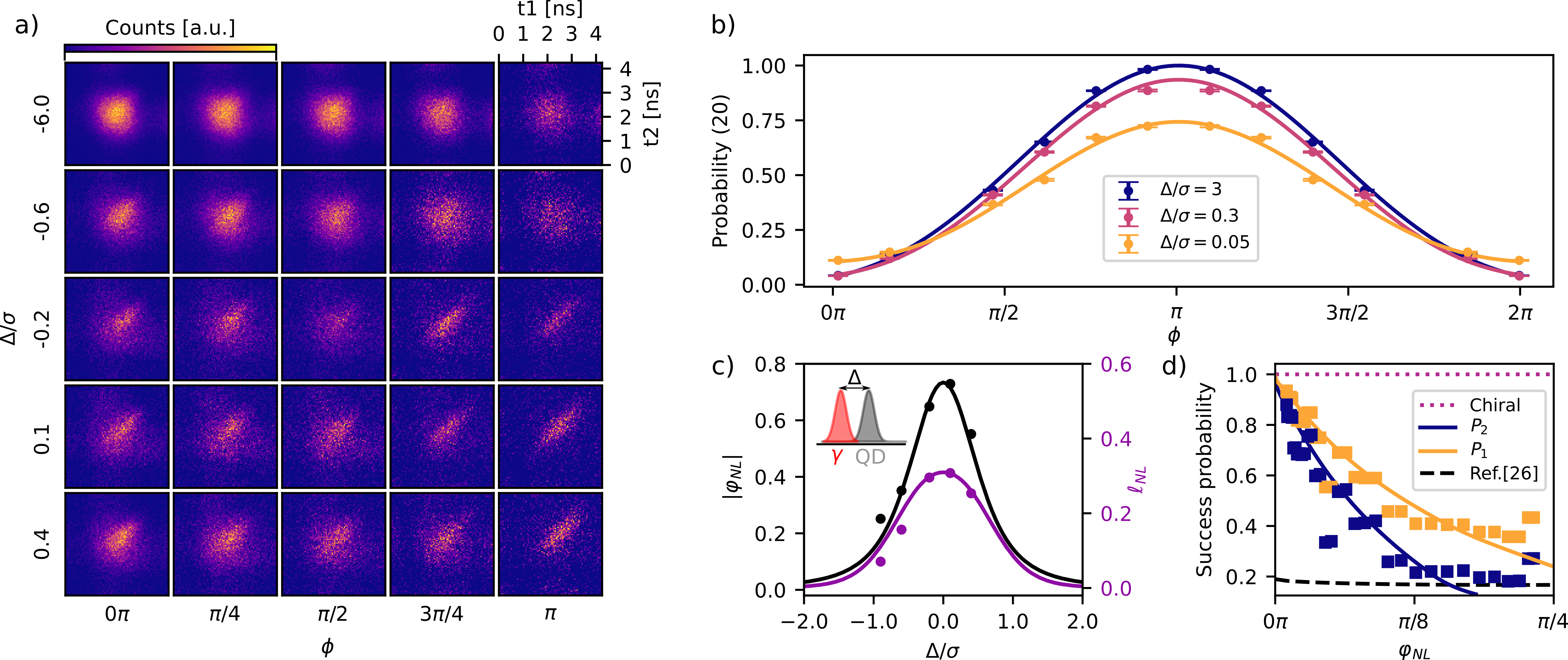}
    \caption{\textbf{Programmability of the linear and nonlinear photonic operations.}
    a) Experimental normalised joint temporal intensity data obtained via time-resolved correlation measurements of the output photons.
    The different columns correspond to the different settings of the linear phase of the interferometer, while the different rows correspond to the different QD detunings in order to control the strength of the nonlinearity.
    All measurements are taken with a 700 ps long input pulse.
    b) Measurements of interference in the circuit for different nonlinearities.
    The measured probability for the output configuration $\ket{20}$ is shown as a function of the linear phase $\phi$ for different values of the detuning $\Delta$ that controls the strength of the nonlinearity. 
    Error bars represent 1 standard deviation and are calculated assuming Poissonian photon statistics.
    c) Characterisation of the nonlinear phase and scattering probability parameters as a function of detuning $\Delta$.
    Black (purple) data represent the nonlinear phase (scattering probability) as extracted from fitting the measured photon statistics to the nonlinear transformation model described in the main text.
    Inset: representation of the detuning $\Delta$ between the incoming photons (labeled as $\gamma$) and the QD used to control the nonlinearity. 
    d) 
    Comparison of the transmission probability after the scattering process for one-photon (yellow) and two-photon (blue) components as a function of the implemented nonlinear phase $\varphi_\text{NL}$.
    Markers are values calculated from the experimental data, while solid lines are theoretical values modeling the experimentally characterised system parameters. 
    Analogue curves are also shown for the cases where scattering in a chiral or single-sided waveguide is employed (dotted purple line, assuming a unit $\beta$), and for the optimized measurement-based linear optical scheme proposed by Sparrow et al.~\cite{sparrow2018} (dashed black line).
    }
    \label{fig:characterization}
\end{figure*}

\ \\

\noindent\textbf{Programmable linear and nonlinear photonic circuits.} 
Fig.~\ref{fig:setup}a displays the schematics of the photonic circuit realizing programmable nonlinear photonic operations between two tunable linear interferometers. 
The scheme employs temporal degrees of freedom and universal two-mode linear operations as performed with two Mach-Zehnder interferometers implemented via a self-stabilizing time-bin interferometer (see Fig.~\ref{fig:setup}b, and Appendix~\ref{appendix:expedetails} for more details).
The temporal mode encoding allows scattering from the same QD at different times allowing to scale up to multiple modes.
The photon-photon interaction is mediated using an InAs QD, which is embedded in a two-sided GaAs photonic crystal waveguide (PCW, shown in Fig.~\ref{fig:setup}b) to enhance light-matter interactions between the incoming photons and the quantum emitter (coupling strength parameter of $\beta \simeq 88\%$) ~\cite{le2021experimental}.
To achieve this configuration, the photons emerging from the first time-bin interferometer are coupled via a grating coupler~\cite{zhou2018} from free-space and into the GaAs chip.
After the interaction with the QD the photons are outcoupled from the chip and directed back to the time-bin interferometer to perform the second linear operation. 
Finally, after the interferometers, the photons are collected via optical fibers, demultiplexed via fiber beam-splitters, and detected with four superconducting nanowire single-photon detectors (SNSPDs) enabling pseudo-number-resolving photon counting. 
We consider input states with two photons in the first input mode, which can be implemented by sending weak-coherent laser pulses in the initial time-bin corresponding to the first mode and condition on two photon detection events at the SNSPDs. 
The input pulses are created by modulating a narrowband continuous-wave (CW) laser through a lithium niobate intensity-electro-optic-modulator (EOM) driven by an arbitrary waveform generator (AWG) that generates Gaussian pulses of duration 300 to 700 ps at a repetition rate of 38.5 MHz, i.e. slower than twice the temporal distance between the two time-bins in order to populate only the first time-bin mode.
The coherent state is attenuated to $n<0.1$ photons per pulse on average at the QD to suppress noise due to multi-photon components.
In- and out-coupling from the GaAs chip is achieved with an efficiency of approximately 60\%, which can readily be improved further by optimized gratings ~\cite{zhou2018} or by utilizing evanescent coupling to a tapered optical fiber~\cite{zeng2023cryogenic}.
To program the photon-photon nonlinearities, we use two control parameters: i) the spectral detuning $\Delta$ between the QD resonance and the photon frequency, and ii) the temporal bandwidth of the photons relative to the intrinsic linewidth, $\sigma$, of the QD resonance (corresponding to a lifetime $2\pi/\sigma=311~\text{ps}$ for the QD used in this work). 
The spectral detuning i) is tuned by controlling the DC voltage applied to the QD, which shifts its resonance due to the DC Stark effect~\cite{kirsanske2017}.
The bandwidth ii) is adjusted by changing the duration of the modulation applied to the EOM via the AWG.
In both cases, the interaction is strongest when the pulses are resonant with the QD and when the bandwidth of the pulses matches or is narrower than the QD linewidth.
The linear operations are programmed by tuning the reflectivity and relative phase of the self-stabilising temporal interferometers by rotating the polarization in each arm of the interferometer~\cite{appel2022}. 
Further details on the experimental apparatus can be found in Appendix~\ref{appendix:expedetails}.

\ \\

\noindent\textbf{Characterisation of the linear and nonlinear operations.}
First, the QD detuning, pulse width and linear phase are scanned to demonstrate the complete and simultaneous linear and nonlinear programmability of the circuit. 
The interplay between linear and nonlinear processes can be observed from the dynamics of the joint temporal intensity (JTI) of the photons which are scattered in the nonlinear transformation and interfered in the linear circuits.
We probe the dynamics in time-resolved photo-correlation measurements of the photons emerging from the final Mach-Zehnder interferometer after the photons have scattered and interfered.
Fig.~\ref{fig:characterization}a displays the normalised two-photon JTI measured in the $\ket{02}$ configuration (both output photons in the second mode)
for different values of both the linear phase $\phi$ and the nonlinearity control parameter $\Delta$.
It is observed that, while no correlations are present in the absence of the nonlinearities ($\Delta \gg \sigma$), they arise when the QD is brought into resonance leading to a JTI with pronounced correlations along the diagonal in addition to persistent side lobes ~\cite{jeannic2022}. 
When the linear phase is scanned, the side lobes interfere due to their emergence in the contributions to the JTI from the terms where only a single photon is present in each time-bin mode during the scattering process, which we label as the $\ket{11}$ term. 
The diagonal part is mostly due to contributions from the terms with both photons in the same mode, i.e. the $\ket{20}$ and $\ket{02}$ terms, and is thus not affected by $\phi$.
%

%
The quantum interference patterns are also affected by the interplay between the linear and nonlinear parameters. 
Fig.~\ref{fig:characterization}b shows the interference fringe in the output configuration with both photons bunched in the first mode, i.e. $\ket{20}$, obtained when scanning the linear phase $\phi$ and setting the Mach-Zehnder interferometers to balanced beam-splitters ($\theta = \pi/2$).
In the absence of nonlinearities (i.e. $\Delta \gg \sigma$), an interference fringe visibility of $97.1\%$ is observed (dark blue curve in Fig.~\ref{fig:characterization}b, see Appendix~\ref{appendix:analysis} for details on the data analysis), which is reduced by the nonlinear phase and amplitude contributions realized when bringing the QD into resonance.
We describe the nonlinear transformation of the scattering process through a nonlinear phase  parameter $\varphi_\text{NL}$ and the scattering probability $\ell_\text{NL}$ according to
\begin{align}
    \ket{1} &\mapsto \sqrt{\eta(1-\ell_\text{NL})} \text{e}^{i \phi} \ket{1}
    \label{eq:eq1}
    \\
    \label{eq:eq2}
    \ket{2} &\mapsto \sqrt{\eta^2} \text{e}^{i (2\phi +\varphi_\text{NL})} \ket{2},
\end{align}
where $\phi$ and $\eta$ are the linear phase and transmission parameters, respectively and the response is truncated to two-photon components. 
We note that this description of the scattering process is a simplified parametrization where both a Kerr-like nonlinear phase as well as the finite overlap between the input and output spectra are embedded in a single nonlinear phase parameter $\varphi_\text{NL}$ (see Appendix~\ref{appendix:from_input_output_to_nl} for details).
Nevertheless, this simplified model  captures well the effects of the nonlinearities observable in our device.  
The nonlinear phase parameter $\varphi_\text{NL}$ manifests itself by reducing the visibility in the interference fringes in Fig.~\ref{fig:characterization}b due to the nonlinear contributions between the two-photon terms and the $\ket{11}$ term.
Using this model for the nonlinear transformation, we extract the nonlinear phase parameter $\varphi_\text{NL}$ and nonlinear scattering probability $\ell_\text{NL}$ as a function of the nonlinearity control parameters by modelling the experimental photon statistics data (see Appendix~\ref{appendix:from_input_output_to_nl} for more details on the model and analysis). 
Fig.~\ref{fig:characterization}c shows the results as a function of detuning $\Delta$.
Equivalent results for controlling the nonlinearities by tuning the bandwidth of the photons are shown in Appendix~\ref{appendix:input_output}.
From the photon statistics we can extract the applied nonlinear phase parameters up to $\varphi_{\text{NL}} \simeq \pi/4$, indicating a large range of tunability of the nonlinearities embedded within the universal programmable two-mode linear interferometer.
The finite nonlinear scattering probability $\ell_\text{NL}$ arises due to the scattering process being implemented in a two-sided  waveguide, where scattered photons can also be reflected.
Because the probability of having at least one photon reflected is higher for the $\ket{1}$ than for the $\ket{2}$ term~\cite{le2021experimental}, this process results in a finite success probability, cf.  Fig.~\ref{fig:characterization}d, which decreases when increasing the strength of the nonlinearity.
Notably the success probability for the nonlinear approach is found to significantly outperform that of the measurement-based scheme based on linear optics (dashed black line in Fig.~\ref{fig:characterization}d)~\cite{sparrow2018}.
Furthermore, full deterministic operation can be readily achieved by implementing either one-sided waveguides~\cite{wang2023deterministic} or chiral photon-emitter interaction~\cite{lodahl2017}  (dotted purple line in Fig.~\ref{fig:characterization}d), for which a unity intrinsic scattering success probability  in the full $[0, \pi]$ range is achievable. 

\begin{figure*}
    \centering    
    \includegraphics[width=1\textwidth]{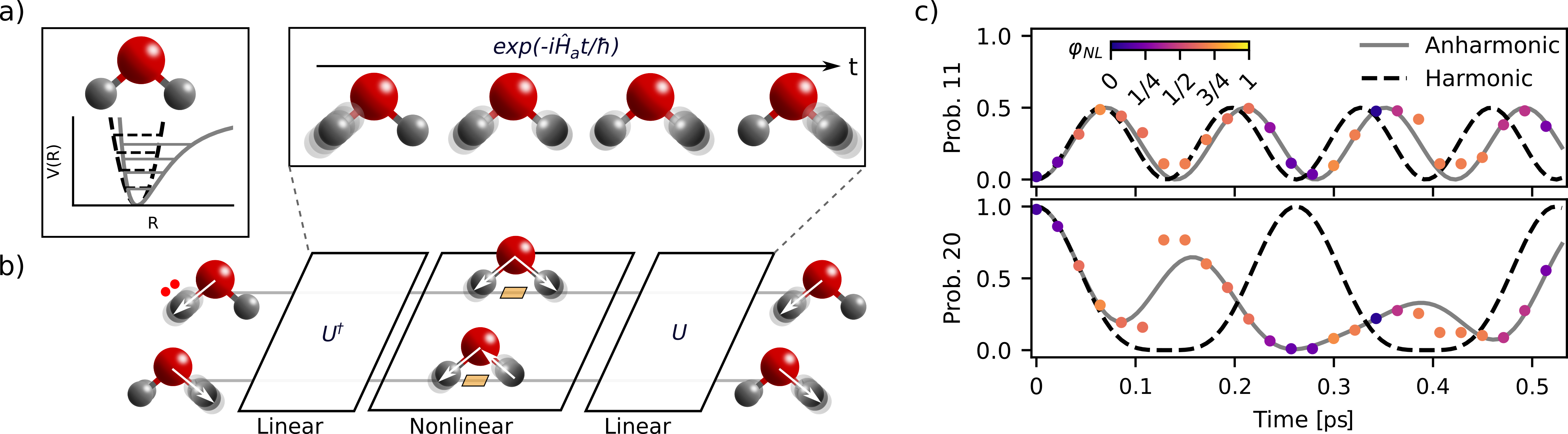}
    \caption{\textbf{Proof-of-concept quantum simulation of vibrational dynamics in $\text{H}_2\text{O}$ with a programmable nonlinear photonic quantum circuit}. 
    a) Illustration of molecular potential of vibrational degrees of freedom that in general is anharmonic (depicted as solid line) i.e. a harmonic approximation (dashed line) is not sufficient. 
    b) The quantum vibrational dynamics  of the molecule is mapped onto a photonic system by associating optical modes to vibrational modes, and single optical excitations (photons) to single vibrational excitations (phonons).
    The associated vibrational dynamics for an anharmonic Hamiltonian $\hat{H}_a$ (upper inset) can be described in the localised basis (depicted on left and right) by converting to the normal basis (depicted in the middle) via a matrix transformation $U$, evolving in the normal basis for a time $t$, and then converting back to the localised basis via $U^\dagger$.
    In the photonic implementation, the linear parts of the circuit perform the basis changes and the harmonic evolution.
    Subsequently, the photonic nonlinear interactions (depicted as yellow boxes) implement the anharmonic part of the evolution. 
    c) Experimental results for the quantum simulation of the anharmonic dynamics of the $\text{H}_2\text{O}$ molecule initialised with two photons in the same localised stretch mode.
    The top and bottom panels report the output occupancy for the configurations where the excitations emerge in separate or in the same localised stretch mode, respectively. 
    The data are fitted to the theoretical model (solid lines), and compared also to the predictions using the harmonic approximation (dashed lines).
    The data markers are color-coded according to the strength of the nonlinear phase (scale bar as inset) associated with each evolution time step.
    }
    \label{fig:water_sim}
\end{figure*}

\ \\
\noindent\textbf{Photonic quantum simulation of anharmonic molecular dynamics.}
In the following, we illustrate the usability of the programamble nonlinear photonic circuits for one exemplary application of quantum simulation of the anharmonic molecular dynamics. 
First proposed in Ref.~\cite{sparrow2018}, this class of quantum simulation methods exploits the equivalence between phonons evolving in the vibrational modes of a molecule and photons evolving in the optical modes of a photonic circuit, as depicted in Fig.~\ref{fig:water_sim}a. 
In fact, because phonons and photons are both bosonic particles, the quantum dynamics of vibrations in a molecule evolving under a transformation $U$ will be described by the same input-output statistics as a photonic system evolving through a circuit implementing an equivalent optical transformation $U$.
In such a mapping, the $N = 3M-6$~\footnote{The vibrational modes are $3M-5$ for linear molecules.} vibrational modes of an $M$-atom  molecule each correspond to an optical mode in the photon circuit. 
In the harmonic approximation, the system is in general represented by $N$ coupled harmonic oscillators, with couplings between the vibrational modes that correspond to linear operations between the associated optical modes.
The evolution of molecular quantum dynamics in the harmonic approximation can thus be simply implemented via a linear optical system: input photons are prepared in the same state as the initial phononic state of the molecule, and the unitary transformation $U$ is implemented by programming the linear circuit. 
The output state is then analyzed via photon-detection~\cite{sparrow2018}.
However, the potential energy landscapes for most molecular systems of practical interest are far from being harmonic, and higher order corrections are required to better capture relevant properties of the system.
In the mapping between vibrational and optical modes, such higher-order corrections correspond to nonlinear coupling terms between the modes, which in the first-order Trotter approximation correspond to single-mode nonlinearities embedded in between two linear optical transformations~\cite{sparrow2018}.
Such a circuit is exactly what we implement in our device, as shown in  Fig.~\ref{fig:setup}, where the capability to tune both linear and nonlinear operations enables direct implementation of the transformations corresponding to the anharmonic dynamics of molecules.
In particular, we simulate the anharmonic vibrational dynamics in the $\text{H}_2 \text{O}$ (water) molecule, which has three vibrational eigenmodes: two high-energy symmetric and asymmetric stretch modes (shown in Fig.~\ref{fig:water_sim}b, with more details in Appendix~\ref{appendix:water}), and a low-energy bending mode, which is uncoupled to the others.
Molecules are often excited in modes localized around a single atom or chemical bond~\cite{gatti2014molecular, likar1988vibrationally} and here the localized modes are represented by stretch modes of the individual left and right H atoms, as shown in Fig.~\ref{fig:water_sim}b.
Because these localized modes are superpositions of the stretch eigenmodes, they are uncoupled from the bending mode, and the system can thus be modeled by restricting to these two high-energy vibrational modes and mapped into the two-mode photonic device.  
In this way, the time evolution of the molecular vibration dynamics of $\text{H}_2 \text{O}$  including its anharmonic potential energy can be simulated by tuning the linear and nonlinear parameters of the quantum photonics circuit.
We consider the $\text{H}_2 \text{O}$ molecule initialized with two excitations (phonons) on the left stretch mode, corresponding to an initial photonic state with two photons on the first optical mode. 
The first linear operation implements a basis transformation of the localised modes into the vibrational eigenmodes of the molecule, for which the evolution for a given time step is then simulated as follows.
From the anharmonic eigenenergies of the vibrational modes, whose values are reported in Appendix~\ref{appendix:water}, we calculate the phases accumulated during the simulated evolution time by the configurations with one and two excitations on each eigenmode, up to simulated evolution times of 0.5 ps (see Fig.~\ref{fig:water_sim}b).
The relative phases between configurations with a different number of excitations in the same mode determine the nonlinear phase $\varphi_{\text{NL}}$ to be applied, while the relative phases between configurations with the same number of excitations correspond to the relative linear phase $\phi$ between the modes~\cite{sparrow2018}.
We then program the detuning $\Delta$ to match the required nonlinear phase according to the relation characterised in Fig.~\ref{fig:characterization}c, while the linear phase $\phi$ is programmed by reconfiguring the linear time-bin interferometers.
The second interferometer transforms the eigenmodes back to the localised modes, and photon counting at the output then provides the statistics for the excitations in the localised modes after the evolution. 
The results are shown in Fig.~\ref{fig:water_sim}c, where we report the probabilities of the phonons in the final molecular state occupying different local modes (top panel) or being bunched in the same mode as the input local mode (bottom panel).  
The results closely follow the theoretical evolution of the anharmonic model (solid line), which differs significantly from the predictions within the harmonic approximation (dashed line).
The applied nonlinear phase shift for each different time step is shown via the color code of the data points. 
We note that the largest differences between experiment and theory are found at the time instances where the strongest nonlinear phase shift is required, which originate from imperfections in the scattering process due to a finite quantum cooperativity. 
Future improvements to the device include reducing minor residual broadening processes to improve quantum cooperativity of the QD photon-emitter interface~\cite{uppu2021} and chiral operation in glide-plane photonic crystal waveguides~\cite{sollner2015} for true deterministic operation.

\ \\

\noindent\textbf{Discussion.} 
We have demonstrated a programmable nonlinear photonic circuit with tunable nonlinearities generated through photon-sensitive scattering with a quantum emitter.
The versatility of this technology was tested by implementing an exemplary application - the analog quantum simulation of vibrational quantum dynamics of a molecule.
The implementation relies on temporal encoding allowing for a highly resource-efficient realization since  multiple nonlinear operations are encoded by reusing the same quantum emitter at different times.
The demonstrated capability to combine linear and nonlinear operations in a photonic device unlocks new opportunities for photonic quantum technologies. 
These include implementing deterministic Bell-state and quantum non-demolition measurements~\cite{witthaut2012photon, ralph2015} and deterministic photonic entangling gates~\cite{Schrinski2022}, tasks that are fundamentally constrained in conventional linear optical approaches posing significant limitations when developing resource-efficient photonic quantum computing architectures~\cite{bartolucci2023, paesani2023} or quantum repeater architectures~\cite{borregaard2020}.
Using nonlinear quantum photonic circuits for further scaling up analog quantum simulations of vibrational dynamics to complex molecular or even biological systems represents an important research direction, potentially opening a route to earlier practical applications for photonic quantum simulators. 
Our resource-efficient approach  is particularly appealing for scaling the nonlinear circuits to target larger molecular systems, and low-loss time-bin linear interferometers with hundreds of modes have already been realised~\cite{madsen2022,motes2014scalable}.
Operating such nonlinear circuits at a sufficiently large scale can potentially realize quantum simulations of complex molecules with a computational advantage over classical simulation techniques.
In this respect, the ability to perform direct nonlinear operations could significantly increase the computational complexity of the simulated problem, thereby reducing the number of photons and hardware performance requirements needed to reach the quantum advantage regime~\cite{spagnolo2023}.
Another promising direction would be to configure the programmable nonlinear quantum photonic circuit as a quantum neural network~\cite{steinbrecher2019} and train the system to synthesize advanced quantum states of light, e.g., multi-photon entangled graph states for photonic quantum computing~\cite{bartolucci2023, paesani2023} or one-way quantum repeaters~\cite{Buterakos2017}, or non-gaussian quantum states such as Schrodinger cat states or GKP states for continuous-variable quantum information processing~\cite{bourassa2021, hastrup2022}.

Our experiment is the first step towards unlocking such opportunities for photonic quantum hardware using programmable nonlinear photonic circuits.

\ \\
\noindent\textbf{Ackowledgements.}
We are grateful to Klaus Mølmer, Bj\"orn Schrinski, Hanna Le Jeannic, and Jacques Carolan for fruitful discussions. 
We acknowledge funding from the Danish National Research Foundation (Center of Excellence “Hy-Q,” grant number DNRF139), the Novo Nordisk Foundation (Challenge project "Solid-Q"), the European Union’s Horizon 2020 research and innovation program under Grant Agreement No. 820445 (project name Quantum Internet Alliance). S.P. acknowledges financial support from the European Union’s Horizon 2020 Marie Skłodowska-Curie grant No. 101063763, from the Villum Fonden research grants No.VIL50326 and No.VIL60743, and from the NNF Quantum Computing Programme.

\ \\
\noindent\textbf{Conflicts of interest}
P.L. is a founder of the company Sparrow Quantum which commercializes single-photon sources. The authors declare no other conflicts of interest.


\bibliography{bib}


\newpage 
\onecolumngrid
\clearpage

\pagenumbering{arabic}
\appendix
\renewcommand{\thesection}
{\Alph{section}}
\renewcommand{\thefigure}{\thesection\arabic{figure}}
\setcounter{figure}{0} 
\renewcommand{\thetable}{\thesection\arabic{table}}
\setcounter{table}{0} 


\section{Experimental setup details}
~\label{appendix:expedetails}
In this section, we provide additional details on the functioning and characterisation of the experimental setup.
A detailed schematic of the setup is shown in Fig.~\ref{fig:s1_setup}a.
It consists of the creation of weak-coherent time-bin states, linear operations implemented through a single double-path time-bin interferometer (TBI), and a light-matter interface with a chip hosting QDs at cryogenic temperatures to implement the nonlinearities.
\subsection{Linear circuits via double-path time-bin interferometers}  	
A single double-path time-bin interferometer (TBI), schematized in Fig.~\ref{fig:s1_setup}a, enables the dual task of manipulating the excitation laser and processing the emitted signals using the two different paths. 
The functioning of the TBI to generate and process time-bin states has already been described and demonstrated in previous works~\cite{appel2022}, and we describe it here for completeness.
The initial time-bin states and linear optical operations over the two time-bin modes are performed in what we call \textit{excitation path} of the interferometer, depicted in red in Fig.~\ref{fig:s1_setup}a. 
The second linear operation is instead performed through the \textit{detection path}, depicted in blue.
The advantage of the double-path configuration for the interferometer is to obtain self-stabilization of random phase fluctuations in the interferometer, allowing us to avoid the need for active phase stabilization in the temporal interferometer.

\subsubsection*{Excitation path }
~\label{Appendix:Exde_Expa}
\begin{figure}
    \centering
    \includegraphics{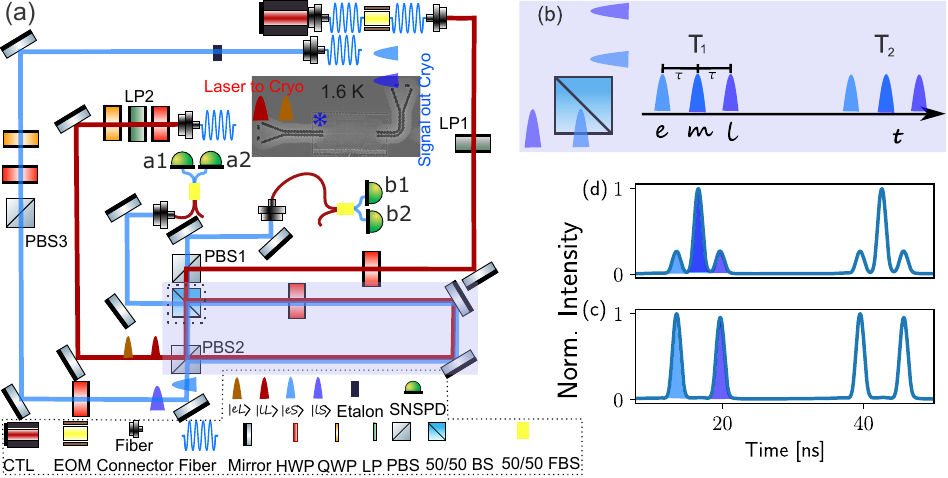}
    \caption{
    (a) Complete schematic of the experimental setup. 
    A free space double-pass time-bin interferometer (TBI) contains the excitation path sketched in red line and the detection path in blue. 
    The main body of the TBI (shaded in light blue) comprises an asymmetric Mach-Zehnder interferometer composed of a polarized beam splitter (PBS2), a 50/50 beam splitter (BS), a pair of mirrors and a half-waveplate (HWP) in the long path. 
     The excitation path prepares the respective early and late excitation pulses depicted in orange and red pulses. 
     The pulses are then coupled to a photonic crystal waveguide embedding QDs placed at a cryogenic temperature of 1.6K. 
     After the nonlinear scattering at the QD, the output time-bin pulses  the early($\ket{eS}$)and late signal($\ket{lS}$) pulses are sent to the detection path of the TBI.
     The resulting time-bin states are illustrated in (b) for three cases: i) the ($\ket{eS}$) transmits the PBS2  to the short path and arrives at the BS denoted as $\ket{ee}$, corresponding to an early ($e$) time-bin.
     ii): The($\ket{lS}$) reflects into the long path and arrives at the BS denoted as $\ket{ll}$, corresponding to a late ($l$) time-bin.
      iii) The ($\ket{eS}$) propagates in the long path while the ($\ket{lS}$) to the short path, arriving at the BS spontaneously denoted as $\ket{el}$ and $\ket{le}$ and corresponding to the middle ($m$) time-bin event. 
      The outputs of the TBI are measured using superconducting nanowire single-photon detectors (SNSPDs).
       (c) and (d) report the experimental histogram registered in the detector $a1$ for a linear phase of $0$ and $\pi$, showing a minimum and maximum, respectively, of the interference in the $m$ state between the early and late time-bins, while no interference in the $e$ and $l$ states.}
    \label{fig:s1_setup}
\end{figure}

To generate the initial weak-coherent time-bin states, we start with a narrow bandwidth continuous-wave laser (Toptica CTL) which is then temporally modulated via a lithium niobate intensity electro-optical modulator (EOM) driven by an arbitrary waveform generator (AWG).
The AWG generates RF driving pulses with a Gaussian waveform at a repetition rate of 38.5 MHz.
These signals are sent to the EOM to modulate the laser light, resulting in the generation of a train of optical Gaussian laser pulses with a separation of 26 ns, as illustrated in Fig.~\ref{fig:s1_setup}b. 
The pulses are attenuated to create weak-coherent states so that, when reaching the QD, the average photon number per pulse is $\langle n \rangle<0.1$.
After passing through a linear polarizer (LP1) to purify the polarization and an half-wave plate (HWP) to prepare the light in horizontal (H) polarization, the pulses are injected into the excitation path of the TBI. 
This is done by first passing the pulses through a polarizing beam splitter (PBS1), which is used to separate the input light into the excitation path from the output light from the detection path of the TBI. 
In the excitation path, the evolution of the states of light can be described as follows.
The electric field at the input is given by the Jones vector
\begin{equation}
    E_L= \varphi(t)
    \begin{pmatrix}
    1 \\
    0 
    \end{pmatrix}	
\label{Eq_ep1}
\end{equation}
where $\varphi(t)$ is the laser envelope function and $(10)$ and $(01)$ are vectors representing the orthogonal basis of horizontal (H) and vertical (V) polarization.
The TBI can be seen as an asymmetric Mach-Zehnder interferometer, with a first 50/50 beam-splitter (BS) dividing the input pulses into a short arm and a long arm with a 1.1 m length difference corresponding to a 3 ns delay between the two arms. 
The pulse propagating through the short path and reflecting at a second PBS (PBS2) generates an early excitation pulse $\ket{eL}$.
The pulse propagating through the long part is first rotated in polarization into V by an HWP at $45^\circ$ and then transmitted at PBS2 to generate the late excitation pulse $\ket{lL}$.
The time-bins $\ket{eL}$ and $\ket{lL}$ represent the two optical temporal modes we use to encode quantum information of the weak-coherent state.
The respective electric fields are given by
\begin{equation}
    E_{eL}= \sqrt{T_{short}}\varphi(t-t_{short})
    \begin{pmatrix} 1 \\ 0  \end{pmatrix} 	
\label{Eq_ep2}
\end{equation}
\begin{equation}
    E_{lL}= \sqrt{T_{long}} e^ {-i\theta}\varphi(t-t_{long})
    \begin{pmatrix} 0 \\ 1  \end{pmatrix} 	
\label{Eq_ep3}
\end{equation}
where $\theta$ is a random phase difference between the long-path and the short-path, which will be compensated when transmitting back through the TBI in the detection path.
The parameters $T_{short}$ and $T_{long}$ represent the transmission efficiency over the short and long arm, respectively, while $t_{long}$ and $t_{short}$ represent their propagation time.

A controllable linear phase between the $\ket{eL}$ and the $\ket{lL}$ pulses is implemented by a quarter waveplate (QWP) and a linear polarizer  (LP2) on the excitation path after PBS2.
The QWP converts the linearly-polarized pulses into circular and LP2 selects the components with the desired relative linear phase $\phi$ between $\ket{eL}$ and the $\ket{lL}$. 
Consequentially, an HWP is installed after LP2 to tune the polarization of the output lights to optimize coupling efficiency into a polarization-maintaining fiber and to the chip hosting the QD. 
Taking into account all these polarization operations, the electric field amplitudes and intensities for each time-bin mode at the output of the excitation path are given by:

\begin{equation}
    E_{out}^{eL}= \left(\frac{1+i}{2} \right)\sqrt{T_{short}}\varphi(t-t_{short})\left(\sin\left( \phi\right) + i\sin\left( \phi-2\theta_{QWP}\right) \right)  
\label{Eq_ep4}
\end{equation}
\begin{equation}
    E_{out}^{lL}= e^{i\theta}\left(\frac{1+i}{2} \right)\sqrt{T_{short}}\varphi(t-t_{short})\left(\cos\left( \phi\right) - i\cos\left( \phi-2\theta_{QWP}\right) \right)  
\label{Eq_ep5}
\end{equation}
\begin{equation}
    I_{out}^{eL}= \abs{E_{out}^{eL}}^2= \frac{T_{short}}{4}\abs{\varphi(t-t_{short})}^2\left(2-\cos\left(2 \phi\right) -i\cos\left( 2\phi-4\theta_{QWP}\right) \right) 
\label{Eq_ep6}
\end{equation}
\begin{equation}
    I_{out}^{lL}= \abs{E_{out}^{lL}}^2= \frac{T_{short}}{4}\abs{\varphi(t-t_{short})}^2\left(2+\cos\left(2 \phi\right) +i\cos\left( 2\phi-4\theta_{QWP}\right) \right).
\label{Eq_ep7}
\end{equation}
To balance the amplitudes of the $\ket{eL}$  and $\ket{lL}$ modes, the QWP is set to $\pi/4$. 
Experimentally, for each $\phi$ the QWP and HWP are rotated to maximize the total transmission.
This is done practically by blocking one arm of the TBI and minimize the power in the fiber until fully suppressing the power by scanning the QWP. 
We then rotate the QWP by $\pi$/4 and tune the HWP to maximize the power in the fiber. 
Repeating this process for every implemented $\phi$, the output excitation field intensity of the $\ket{eL}$ and the $\ket{lL}$ is independent of $\phi$.
The total relative phase between the two generated time-bin modes at the output of the excitation path is then calculated as
\begin{equation}
\begin{split}
    \delta\theta & = \arg {\left(E_{out}^{el}\right)} - \arg {\left(E_{out}^{lL} \right)} \\
     &= \theta + \arg{ \left(\frac{cos\left(\phi\right)-i\cos\left( \phi-2\theta_{QWP}\right) }{\sin\left( \phi\right) + i\sin\left( \phi-2\theta_{QWP}\right)}\right) } \\  
     &= \theta + 2\phi-\pi/2.
\end{split}
\label{Eq_ep8}
\end{equation}

\subsubsection*{ Detection path}

After scattering through the QD device, the early and late time-bin pulses, which we now label as $\ket{eS}$ and $\ket{lS}$ to indicate states after the scattering, are collected by a single-mode fiber and coupled to the detection path of the TBI for the second linear optical operation. 
This is shown in Fig.~\ref{fig:s1_setup}a, where the detection path is depicted in blue.
A QWP and HWP are installed in front of a PBS (PBS3) at the entrance of the detection path to fix the polarization of the signals and maximize the transmission.
Afterward, an HWP is used to control the splitting ratio of the signals divided into the short arm and the long arm of the TBI by PBS2. 
Once the early and late pulses are split, propagated through the short and long arms, and interfered at the 50/50 BS, three time bins emerge, as shown in Fig.~\ref{fig:s1_setup}b.
An early-early time bin (labeled $e$) where the state $\ket{eS}$ undergoes the short path and thus does not interfere with components from $\ket{lS}$.
Similarly, a late-late time bin (labeled $l$) where $\ket{lS}$ undergoes the long path and thus does not interfere with components from $\ket{eS}$.
The middle time bin (labeled $m$) instead represents a component where the state $\ket{eS}$ undergoes the long path and interferes at the BS with the component from $\ket{lS}$ that undergoes the short path. 
After this transformation in the TBI, the two outputs from the BS are measured via pseudo-number-resolving photo-detection by using one balanced fiber beam-splitter per mode and four SNSPDs. 
The output electric fields of the signals after the TBI transformation and at the detectors are described as  
\begin{equation}
\begin{pmatrix} E_{a1}^e \\ E_{b1}^e \\ E_{a1}^m \\ E_{b1}^m \\ E_{a1}^l \\ E_{b1}^l \\ \end{pmatrix}
= 
   \begin{bmatrix} \eta_{Sa1} & 0 \\  
   \eta_{Sb1}e^{i\theta_1} & 0 \\ 
   \eta_{La1} e^{i(\theta_2+\theta')}  & \eta_{Sa1}\\  
   \eta_{Sb1}e^{i\theta'} & \eta_{Sb1}e^{i\theta_1} \\ 
   0 &  \eta_{La1}e^{i\theta_2} \\
   0  & \eta_{Lb1}   \\
   \end{bmatrix}
\begin{pmatrix} E_{out}^{eL}&  E_{out}^{eL} \\  \end{pmatrix}.   	
\label{Eq_ep9}
\end{equation}
Here, $a1$ and $b1$ label the spatial modes associated to each SNSPD, as shown in Fig.~\ref{fig:s1_setup}a. 
The factors $\eta_{Sa1}$ and $\eta_{Sb1}$ indicate the combined efficiency of the short arm and of the detectors $a1$ and $b1$, respectively, and $\eta_{La1}$ and $\eta_{Lb1}$ are analogous parameters for the long arm.
The phases $\theta_1$ and $\theta_2$ indicate the phase difference for the relevant events of the transmitted to reflected and the reflected to transmitted at the final BS. 
For a standard 50/50 BS, $\theta_1$ = $\theta_2$ = $\pi$/2. 
$\theta$' represents the phase difference between the long arm and the short arm of the detection path within the TBI. 
In the absence of nonlinearities (e.g. when the QD is far detuned), for the central $m$ time bin the intensities at two detectors in different arms are given by 

\begin{equation}
    I_{a1}^m = \abs{E_{a1}^m}^2= \eta_{Sa1}^2 +  \eta_{La1}^2 -2\eta_{Sa1}\eta_{La1}\sin{(\theta_2+\theta'-\theta-2\phi)}
    \label{Eq_ep10}
\end{equation}
\begin{equation}
    I_{b1}^m= \abs{E_{b1}^m}^2= \eta_{Sb1}^2 +  \eta_{La1}^2 +2\eta_{Sa1}\eta_{La1}\sin{(\theta_1-\theta'+\theta+2\phi)}
    \label{Eq_ep11}
\end{equation}
which present interference fringes given by
\begin{equation}
    \frac{I_{a1}^m-I_{b1}^m}{I_{a1}^m+I_{b1}} = - C \cos{(2\phi+\theta-\theta'-\frac{\pi}{2})}
    \label{Eq_ep12}
\end{equation}
where 
\begin{equation}
    C =\frac{2 (\eta_{Sa1}\eta_{La1} + \eta_{Sb1}\eta_{Lb1})}{\eta_{Sb1}^2+\eta_{La1}^2+\eta_{Lb1}^2+\eta_{Sa1}^2} 
\end{equation}
is due to possible efficiency imbalances.
These linear optical operations are thus equivalent to the time-bin operations schematised in Fig.~\ref{fig:setup} of the main text.

The equations above represent interference fringes appearing as a function of the controllable linear phase $\phi$, as well as of $\theta$ - $\theta$' which is the difference of the relative phase between the short and long arms in the TBI for the excitation and detection paths.
Since random drifts in the TBI affect the excitation path and detection path equally, $\theta$ - $\theta$' remains constant, which represents the self-stabilization of our two-path TBI and allows us to avoid active phase stabilization. 
Note, however, that while $\theta$ - $\theta$' is constant it can be non-zero, requiring initial calibration of the TBI to characterize it.
Such calibration procedure is described in the following section.

\begin{figure}
    \centering \includegraphics{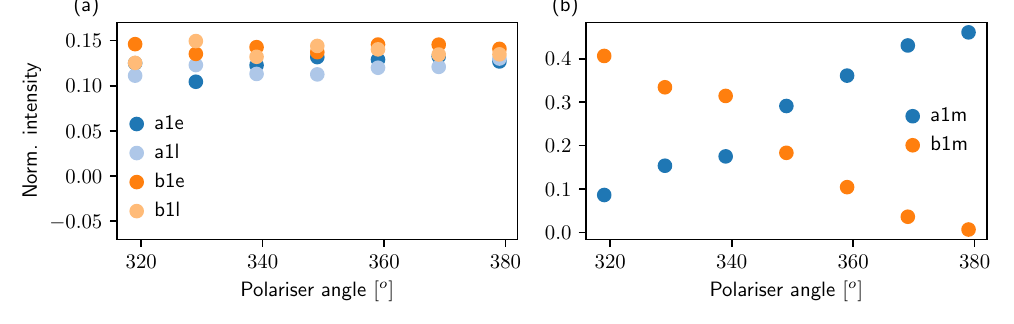}
    \caption{The integrated intensity in the output time-bins of the TBI when verying the linear phase $\phi$. (a) reports the normalised intensity for the $e$ and $l$ detection window, and (b) reports them for the $m$ detection window.}
 \label{fig:S1-TBI_intensity}
 \end{figure}

\begin{figure}
    \centering \includegraphics{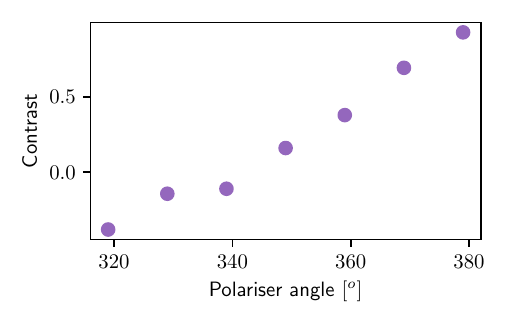}
    \caption{The intensity contrast between the detectors $a1$ and $b1$ for the $m$ output time bin for different values of $\phi$. 
    }
 \label{fig:S1-TBI_contrast}
 \end{figure}

\subsubsection*{ Time-bin interferometer calibration} 
As described in Eq.~\ref{Eq_ep12} in the previous section, interference fringes in the TBI interferometer appear in the central $m$ time-bin as a function of the implemented linear phase $\phi$ up to a constant offset $\theta - \theta'$ given by the difference of the relative phases between the interferometer arms for the excitation and detection paths. 
Calibrating for this offset is equivalent to characterizing the 0 phase for $\phi$.
Experimentally, we do it by scanning $\phi$ to find the phase $\phi_0$ that maximally suppresses the $m$ detection events for the detector $a1$, as shown in Fig.~\ref{fig:s1_setup}c. 
As expected, for such $\phi_0$ the $m$ detection events for the detector $b1$ are maximized, as shown in Fig.~\ref{fig:s1_setup}d. 
We also show in Fig.~\ref{fig:S1-TBI_intensity}a the intensity for the $e$ and $l$ time bins at the output of the TBI when scanning $\phi$ from the calibrated $\phi_0$ with step of 10$^{\circ}$.
As expected, the intensity remains constant as no interference is occurring. 
In Fig.\ref{fig:S1-TBI_intensity}b we instead plot the intensity for the central $m$ time-bin, for which strong interference fringed are instead observed.
In Fig.~\ref{fig:S1-TBI_contrast} we plot the associated intensity contrast between the two detectors (without any background subtraction).
A high contrast of 97.12\%, which corresponds to the interference visibility, is observed at the zero phase $\phi_0 \sim$ 380$^{\circ}$.
\subsection{Nanophotonic device for the nonlinear interface with a quantum dot}

\subsubsection*{Fabrication of the photonic nanostructure}

\begin{figure}
    \centering
    \includegraphics{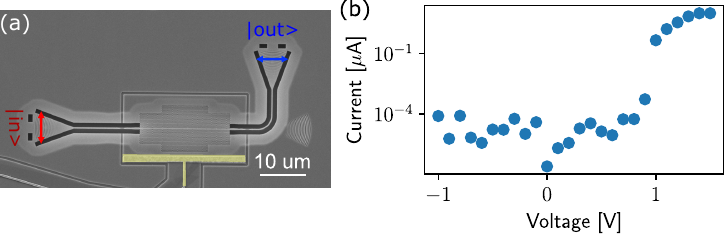}
    \caption{(a) Scanning electron microscope (SEM) image of the photonic crystal waveguide device, which entails the quantum dot implementing the nonlinear light-matter interface. (b) The device response to the applied DC voltage bias across the membrane, which is used to program the strength of the implemented nonlinearities.}
 \label{fig:S1-sEM-IV}
 \end{figure}

An image of the designed and fabricated integrated nanodevice with a photonic crystal waveguide (PCW) and grating couplers is shown in Fig.\ref{fig:S1-sEM-IV}a. 
The device is fabricated out of a 180nm thick membrane made up of silicon-doped n-layer, intrinsic layer, and carbon-doped p-layer from the bottom to top. 
Quantum dots are grown in the middle of the membrane. 
A p-electrode of Cr/Au is deposited on the surface of the membrane, as shown Fig.~\ref{fig:S1-sEM-IV}a with a false yellow color pattern, while the n-electrode of  Ni/Ge/Au/Cr/Au is deposited on the buried n-layer. 
The p-n electrodes create access to an external voltage source, allowing the tuning of the strength of the non-linearity by tuning the QD resonance by an offset $\Delta$ (see main text) via DC Stark tuning. 
In addition, the externally applied electric field enables suppression of the charge noise. 
In Fig.~\ref{fig:S1-sEM-IV}b we report the characterised I-V curve of the sample.
The bias required for tuning the QD is between 1.2V to 1.3V where the corresponding current is sufficiently low. 
As the bias voltage increases above 1.5V, the current exponentially rises to a few $\mu$A.
A pair of grating couplers with orthogonal orientations are designed to couple the photons in and out of the PCW with high efficiency. 
These couplers are shallowly etched using reactive-ion etching (RIE) to obtain high photon collection efficiency ~\cite{uppu2020scalable}. 
The PCW is etched deeply after the grating couplers using ICP-RIE. 
Finally, wet etching employs 5\% hydrofluoric acid to release the membrane ~\cite{midolo2015soft}. 

\subsubsection*{Resonant transmission measurements}
Resonant transmission measurements are employed to characterize QDs that are efficiently coupled and Purcell-enhanced by the waveguide mode. 
The transmission intensity for CW light is given by ~\cite{javadi2015single}

\begin{equation}
    T=1-\frac{\beta(2-\beta)}{(1+2\frac{\Gamma_d}{\Gamma})(1+\frac{nr}{n_c})}
\end{equation}
where the $\beta$ quantifies the proportion of the light couples to the waveguide mode, defined as $\beta = \frac{\gamma_{wg}}{\gamma_{wg}+\gamma_{loss}}$, $\gamma_{wg}$ is the QD decay rate in the waveguide, and $\gamma_{loss}$ is the decay rate of residual emission not captured by the waveguide. 

\begin{figure}
    \centering
    \includegraphics{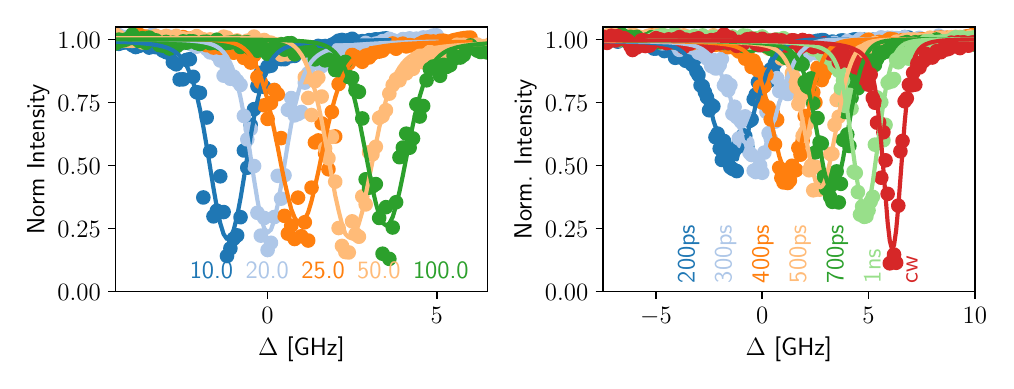}
    \caption{Resonant transmission intensity when (a) increasing the excitation power for a continuous wavelength (CW) laser, and (b) changing the pulse duration at fixed power when sending pulses. 
    In all plots, the measured data points are dots while the solid curves are fits using a Voigt models. 
    To display each transmission spectrum clearly, plots are sequentially shifted on the x-axis.}
    \label{fig:S_RT}
\end{figure}

$\Gamma$ is obtained as $\frac{\gamma_{tot}}{2\pi}$  
which is the natural Fourier transform-limited linewidth at full width at half maximum (FWHM). 
$\Gamma_d$ is the pure dephasing induced by phonons, causing homogenous broadening in the QD spectrum. 
The parameter $n_r$ represents the mean photon number within the QD lifetime, while $n_c$ represents the critical photon number, given by $\frac{1+2(\frac{\Gamma_d}{\Gamma})}{4\beta^2}$. 
The fraction $\frac{n_r}{n_c}$ quantifies the effective saturation of the QD transition. 
To evaluate how well the QD couples to the waveguide mode, weak excitation of $n_r \ll n_c$ is required. 
Such a weak excitation is also required when implementing strong photon-photon nonlinearities mediated by the QD. 
Transmission profiles with different input powers are measured via connecting the output to a SNSPD, and are shown in Fig.~\ref{fig:S_RT}a. 
The reported optical powers are read from an external power meter before coupling into the chip, and are thus not the effective powers reaching the QD. 
The resonant dips do not decrease as the input power increases from 10 nW to 100 nW. 
This implies maximal power applied in the measurement remains in the weak excitation regime in this power range. 
All the data in the main text and the rest of the supplementary is collected with an input power of 50 nW.

To determine $\beta$ and residual spectral diffusion $\sigma_{sd}$, a Voigt fit model is applied to analyze the transmission spectrum. 
The Lorentzian lineshape, characterized by  $\Gamma_{FWHM}= (\Gamma + \Gamma_d) \sqrt{1+\frac{n_r}{n_c}}$, is convolved with a Gaussian lineshape arising from spectral diffusion induced by charge noise. 
$\Gamma$ and $\Gamma_d$ are obtained from the resonant fluorescence characterization (see next section). 
The fittings give $\beta \geq  88 \%$ and the spectral diffusions shown in table \ref{tab:RT_lw} for different laser powers. 
As expected, the spectral diffusion is power-independent in the weak excitation regime. 
Pulsed laser light instead of CW laser light is used when operating the device with time-bin states.
The transmission intensity from different pulse durations is seen in Fig.~\ref{fig:S_RT}b. 
The resonant dip decreases and the spectrum broadens as the pulse duration decreases. 
\begin{table}[]
    \centering
    \begin{tabular}{c|c|c|c|c|c}
         power ($\mu$W) & 10 & 20 & 25  & 50 & 100 \\ \hline
         $\sigma_{sd}$ (MHz) & 124.01$\pm$ 39.18 & 149.88 $\pm$ 10.45 &178.99$\pm$ 11.46 &  134.30 $\pm$ 12.99 & 116.15 $\pm$ 21.86  \\ \hline    
    \end{tabular}
    \caption{Spectral diffusion $\sigma_{sd}$ for different input laser powers, estimated via fitting the resonant transmission spectra to a Voigt model.}
    \label{tab:RT_lw}
\end{table}

\subsubsection*{Resonant fluorescence measurements}
~\label{RF}
\begin{figure}
    \centering
    \includegraphics{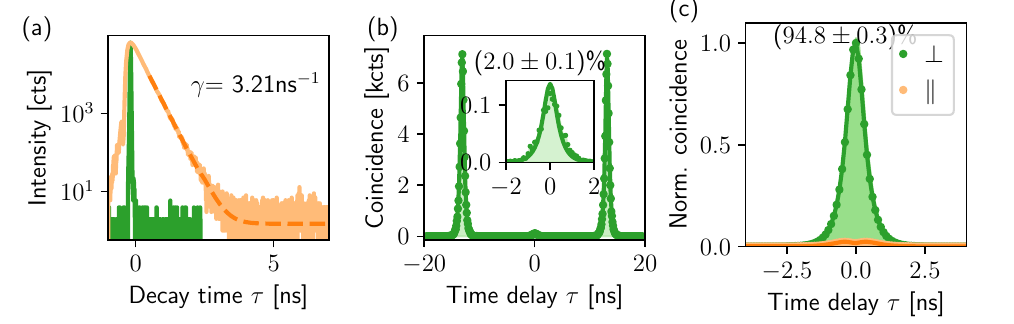}
    \caption{QD characterization under pulsed resonant fluorescence. 
    (a) The measured time-resolved intensity is reported as orange curve and fitted to a single exponential (dashed orange line) with the deconvolution of the instrumental response function (shown as green curve). 
    (b) Characterisation of the single photon purity via the Hanbury Brown and Twiss (HBT) method.
     Insert: zoom on the the zero-time delay area.
     A $g^2(0)$ value of of 2 \% at $\pi$ pulse excitation is observed. 
     (c) Single photon indistinguishability characterisation performed through an un-balanced  Hong-Ou-Mandel interferometer. The co-polarization configuration coincidence curve is shown as orange while the cross-polarization coincidence correlation is plotted in green.}
    \label{fig:S_life_g2_HOM}
\end{figure}

Resonant fluorescence is employed to further characterise the QD device by using an excitation laser focused directly on the QD \cite{uppu2020scalable} using a coherent Mira operating at a high repetition rate of 76 MHz. 
The laser is mode-locked to the resonant frequency of the QD and reshaped using a 4f stretcher to excite the QD. 
By further controlling the excitation laser's polarization, a high signal-to-noise ratio is achieved, facilitating high-fidelity characterization. 
Time-resolved intensity measurements yield a QD decay rate of 3.21 GHz (see Fig.~\ref{fig:S_life_g2_HOM}a), corresponding to a lifetime of 311 ps, indicating a Purcell factor of approximately 3 \cite{RevModPhys.87.347}. 
This corresponds to a Fourier-transform limited linewidth of 510 MHz.

Hanbury-Brown-Twiss (HBT) measurements reveal a $g^2(0)$ value of 2\%, signifying that the QD produces highly pure single photons. 
An unbalanced Hong-Ou-Mandel interferometer, with a long arm delayed by exactly 13.15 ns, matching the repetition rate of the coherent Mira, is employed to interfere consecutively emitted single photons, achieving an indistinguishability of 94.8\% with calibration to account for setup imperfections. 
The partially distinguishable emitted photons are attributed to pure dephasing induced by acoustic phonon vibrations ~\cite{PhysRevLett.120.257401}.

\section{Modeling of the photon scattering process}

~\label{appendix:input_output}

In this Appendix we describe the model  used to describe the nonlinearities that result from the scattering of photons through the QD in a two-way PCW. 
We start describing a general model using the input-output formalism, and then proceed describing a simplified parametrization that can be used to capture the phenomenology observed in our device.

\subsection{Review of input-output theory}

The Hamiltonian governing the interaction between photon pulses and a two-level system comprises three components i) an optical term, ii) an 'atomic' term, and iii) the interaction between photons and the atom~\cite{fan2010}.
The optical component of the Hamiltonian is given by
\begin{equation}
    H_0=\int_{-\infty}^{\infty}\dd\omega \omega  \left( r^\dagger_\omega r_\omega +l^\dagger_\omega l_\omega\right)
    \label{eq_H0}
\end{equation}
where $r,l$ are the operators for the respective right and left propagating photons with frequency of  $\omega$. 
The rest of the system Hamiltonian under the rotation wave approximation scheme \cite{fan2010} is given by
\begin{equation}
        H_1=\frac{1}{2}\Omega \sigma_z+\frac{V}{\sqrt{v_g}}\int_{-\infty}^{\infty}\dd \omega  \left[\sigma_+\left(r_\omega +l_\omega\right)+\sigma_-\left(r_\omega^\dagger+l_\omega^\dagger \right)\right]
\label{eq_Hi}
\end{equation}
$\Omega$ is the transition frequency of the QD (two-level system), $\sigma_z = 2\sigma^+\sigma^--1$, $\sigma^+$($\sigma^-$) is the atomic transition operator, $V, V_g$ denotes the QD waveguide coupling constant and the photon group velocity respectively. 
The input photons are coupled into the waveguide to interact with the QD using the left grating coupler, see Fig.~\ref{fig:S1-sEM-IV}a.  Afterward, resulting photons propagate in both left and right directions. Our detectors are positioned on the right side. The output state of the one-photon state is expressed as,
\begin{equation}
    \ev{r_{\text{out}}(p)r_{\text{in}}^\dagger(k)}{0}=\overline{t}_k\delta(k-p)
    \label{eq_1p}
\end{equation}
where $\overline{t}_k = \frac{1}{2}\left(\frac{(k-\Omega)-i/\tau}{(k-\Omega)+i/\tau}+1\right)$ denotes the one-photon transmission coefficient. Here, $k$ ($p$) is the frequency of the input (output) photon, and $\tau$ is the lifetime/2 of the QD. The two-photon state is given,
\begin{equation}
\begin{split}
        \ev{r_{\text{out}}(p_1)r_{\text{out}}(p_2)r_{\text{in}}^\dagger(k_1)r_{\text{in}}^\dagger(k_2)}{0}=\\
        \overline{t}_{k_1}\overline{t}_{k_2}\left[\delta(k_1-p_1)\delta(k_2-p_2)+\delta(k_1-p_2)\delta(k_2-p_1)\right]+\\
        \frac{i}{2\pi\sqrt{\tau}}\delta(k_1+k_2-p_1-p_2)s_{p_1}s_{p_2}(s_{k_1}+s_{k_2}),
\end{split}
\label{eq_2p}
\end{equation}
where $s_k=\frac{\sqrt{2/\tau}}{(k-\Omega)+i/\tau}$. The transmitted two-photon state consists of two components. The first term conserves the energy of input-output photons pairwise. Consequently, this is comparable to the two uncorrelated photons and the transmission efficiency is the production of the two individual single-photon transmissions. The second term conserves the total energy of the system and creates a correlation between the two photons. This term introduces a non-linear effect. For more details on the derivations above see \cite{fan2010}.

The equations above are for monochromatic light. In the experiment, input photons are temporally shaped with approximately a Gaussian profile expressed as
\begin{equation}
    \phi(\omega_k, \sigma) = \frac{1}{\sqrt[4]{2\pi\sigma^2}}e^{-\frac{(\omega_k - \omega_0)^2}{4\sigma^2}}
\label{eq_input}
\end{equation}
Where $\omega_k = (k-\Omega) \tau$. Similarly, the pulse width $\sigma$ of the Gaussian spectrum is normalized to $\tau$ as well. Accordingly, the one-photon state \ref{eq_1p} with a Gaussian pulse input reads as
\begin{equation}
   \begin{split}
  &\ev{r_{\text{out}}(p)r_{\text{in}}^\dagger(k)}{0} \\
  & = \overline{t}_k\phi(\omega_k, \sigma)\\
   &=\frac{1}{2}\left(\frac{(k-\Omega)-i/\tau}{(k-\Omega)+i/\tau}+1\right) \frac{1}{\sqrt[4]{2\pi\sigma^2}}e^{-\frac{(\omega_k - \omega_0)^2}{4\sigma^2}}
   \label{eq_1p_Gaus}  
   \end{split}
\end{equation}
which indicates that the output single-photon wavepacket remains identical to the input. The amplitude is modulated by the transmission coefficient determined by the $\beta$ and the input photon frequency $k$.

\begin{figure}
    \centering
\includegraphics{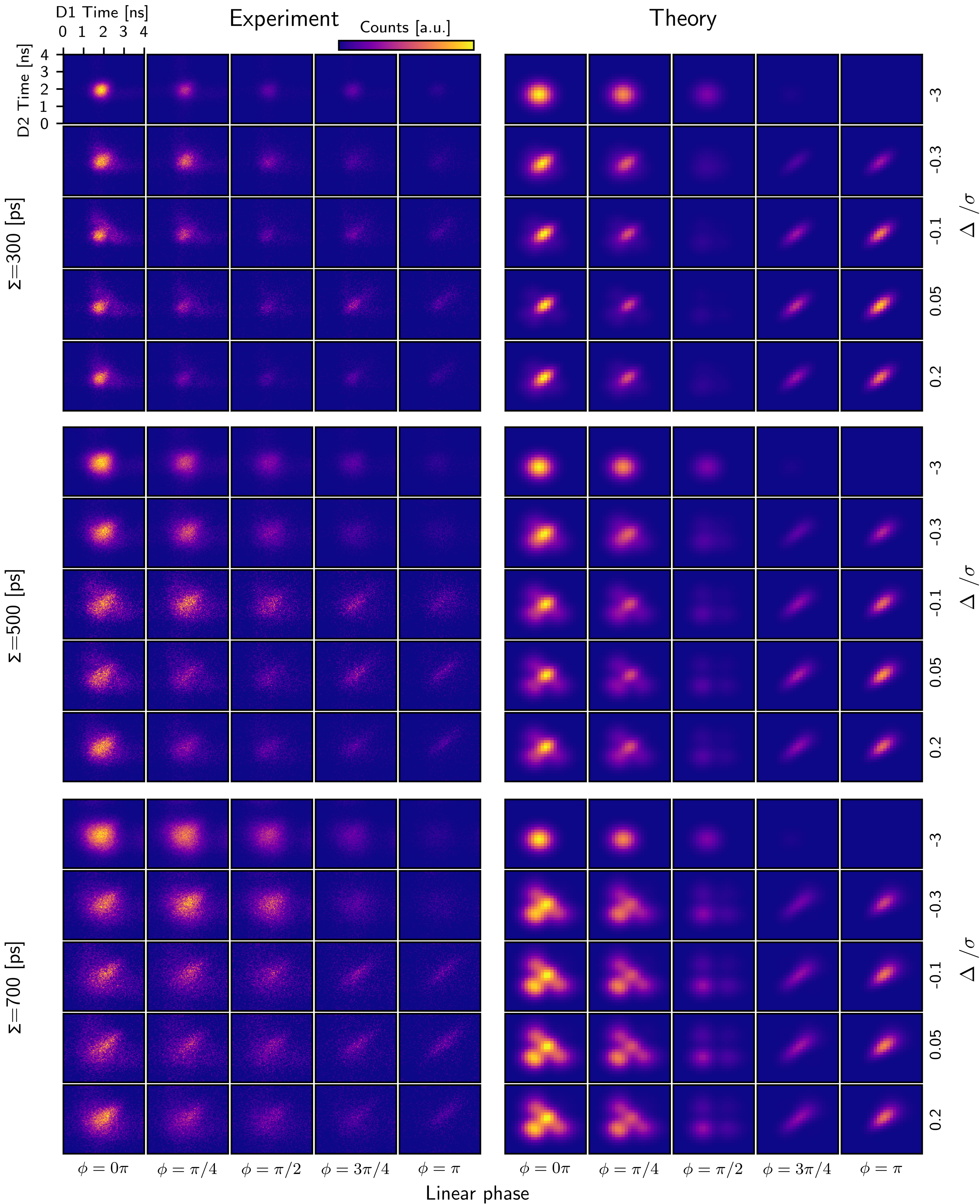}
    \caption{Extensions of the data and analysis shown in Fig.~\ref{fig:characterization}, reporting the experimental (left) and theoretical (right) JTI for various values of the linear phase $\phi$, the relative detuning $\Delta/\sigma$, and the input pulse duration $\Sigma$.}
    \label{fig:butterflyall}
\end{figure}

The two-photon output state is expressed as. 
\begin{equation}
  \begin{split}
   &\ev{r_{\text{out}}(p_1)r_{\text{out}}(p_2)r_{\text{in}}^\dagger(k_1)r_{\text{in}}^\dagger(k_2)}{0} \\
   =&\overline{t}_{k1}\overline{t}_{k2}\phi(\omega_{k1}, \sigma)\phi(\omega_{k2}, \sigma) + \int d \omega_{k1} f_B( \omega_{k1} \phi(\omega_{k1}, \sigma)\phi(\omega_{k2}+\omega_{p2}-\omega_{k1}, \sigma)\\
  =&\overline{t}_{k1}\overline{t}_{k2}\frac{1}{\sqrt{2\pi\sigma^2}}e^{-{\frac{(\omega_{k1} - \omega_0)^2}{4\sigma^2}-\frac{(\omega_{k2} - \omega_0)^2}{4\sigma^2}}}+\\
  &\frac{i}{2\pi} \int_{-\infty}^{\infty} \frac{e^{-\frac{(\omega_{k1} - \omega_0)^2}{4\sigma^2}-\frac{(\omega_{k2}+\omega_{p2} -\omega_{k1} - \omega_0)^2}{4\sigma^2}}}{(i+\omega_{k2})(i+\omega_{p2})\sqrt{2\pi\sigma^2}} \left(\frac{1}{i+\omega_{k1}} + \frac{1}{i+\omega_{k2}+\omega_{p2}-\omega_{k1}}\right) d\omega_{k1}
  \label{eq:input-output}
  \end{split}
\end{equation}
Changing both the input-photon frequency and FWHM has been shown to not only impact the single-photon transmission coefficient but also adjust the correlation observed in two-photon states.
Experimental and theoretical time-resolved joint temporal intensities calculated with this model are reported in Fig.~\ref{fig:butterflyall} as a function of both control parameters we use to tune the strength of the implemented nonlinearities: the QD spectral detuning $\Delta$ and the pulse bandwidth. 

\subsection{From input-output theory to a nonlinear phase}
~\label{appendix:from_input_output_to_nl}
Here we describe the mapping between the more general input-output theory and the simplified model that we use in Eq.~\ref{eq:eq1} and Eq.~\ref{eq:eq2} to represent the nonlinear phenomena that we observe in our device.
As described in the main text, we will truncate our description to one- and two-photon terms. 
As shown in Fig.~\ref{fig:setup}a), in our circuit we start with a state $\ket{20}$, then perform a linear operation composed of a beam-splitter and a linear phase $\phi$.
These linear operations transform the state as:
\begin{equation}
    \begin{split}
      \ket{20}=&\frac{1}{\sqrt{2}}a_0^{\dagger2} \ket{0}\xrightarrow{BS}\frac{1}{2\sqrt{2}}\left(\hat{a}_0^\dagger+\hat{a}_1^\dagger\right)^2\ket{0}=\frac{1}{2\sqrt{2}}\left(\hat{a}_0^{\dagger2}+\hat{a}_1^{\dagger2}+2\hat{a}_0^\dagger a_1^\dagger\right)\ket{0},\\
      \xrightarrow{\phi}&\frac{1}{2\sqrt{2}}\left(\hat{a}_0^{\dagger2}e^{i2\phi}+\hat{a}_1^{\dagger2}+2\hat{a}_0^\dagger \hat{a}_1^\dagger e^{i\phi}\right)\ket{0}.
    \end{split}
\end{equation}

After this initial linear interferometer, the nonlinear operations are performed through scattering with the QD on each individual mode.
An important effect that emerges in the scattering process is the distortion of the spectral wavefunction of the photons, as contained in Eq.~\ref{eq:input-output}. 
In terms of second quantization operators, we describe this effect as a mapping from the old operators to new operators $\hat{a}_0^\dagger \mapsto \hat{a}_0^{\dagger '}$, and we will denote Fock states in these updated modes as for example $\hat{a}_0^{\dagger '} \ket{0} = \ket{1'}$.
Including the finite probability of transmitting the single-photon components, the scattering process can thus be represented as: 

\begin{equation}
   \hat{a}_0^\dagger \hat{a}_1^\dagger \rightarrow \sqrt{\eta^2(1-\ell_\text{NL})^2}\hat{a}_0^{\dagger '} \hat{a}_1^{\dagger '},
   \quad \hat{a}_0^{\dagger 2} \rightarrow \sqrt{\eta^2}\hat{a}_0^{\dagger 2},
   \quad \hat{a}_1^{\dagger 2} \rightarrow \sqrt{\eta^2}\hat{a}_1^{\dagger 2}.
\end{equation}

The circuit then continues by interfering the two modes in a second linear circuit, providing

\begin{equation}
  \begin{split}
    \frac{1}{2\sqrt{2}}&\left(\hat{a}_0^{\dagger2}e^{i2\phi}+\hat{a}_1^{\dagger2}+2\hat{a}_0^\dagger \hat{a}_1^\dagger  e^{i\phi}\right)\ket{0}\\
    \xrightarrow{NL}\frac{1}{2\sqrt{2}}&\left(\sqrt{\eta^2}\hat{a}_0^{\dagger2}e^{i2\phi}+\sqrt{\eta^2}\hat{a}_1^{\dagger2}+
    2\sqrt{\eta^2(1-\ell_\text{NL})^2}\hat{a}_0^{\dagger '} \hat{a}_1^{\dagger '} e^{i\phi}\right)\ket{0}\\
    \xrightarrow{BS}\frac{1}{4\sqrt{2}}&\left[\sqrt{\eta^2}\left(\hat{a}_0^\dagger+\hat{a}_1^\dagger\right)^2e^{i2\phi}+\sqrt{\eta^2}\left(\hat{a}_0^\dagger-\hat{a}_1^\dagger\right)^2+2\sqrt{\eta^2(1-\ell_\text{NL})^2}
    \left(\hat{a}_0^{\dagger '}+\hat{a}_1^{\dagger '}\right)\left(\hat{a}_0^{\dagger '}-\hat{a}_1^{\dagger '}\right)e^{i\phi}\right]\ket{0}\\
    =\frac{1}{4\sqrt{2}}&\left[\sqrt{\eta^2}\left(\hat{a}_0^{\dagger 2}+\hat{a}_1^{\dagger 2}+2\hat{a}_0^\dagger \hat{a}_1^\dagger\right)e^{i2\phi} +\sqrt{\eta^2}
    \left(\hat{a}_0^{\dagger 2}+\hat{a}_1^{\dagger 2}-2\hat{a}_0^\dagger \hat{a}_1^\dagger\right)+2\sqrt{\eta^2(1-\ell_\text{NL})^2}\left(\hat{a}_0^{\dagger ' 2}-\hat{a}_1^{\dagger ' 2}\right)e^{i\phi}\right]\ket{0}.
  \end{split}
\end{equation}

Output statistics of the devices can be readily obtained from this equation. 
For example, the amplitude for the $\ket{20}$ output configuration is calculated as

\begin{equation}
  \begin{split}
    \frac{1}{4\sqrt{2}}\left[\sqrt{\eta^2}\hat{a}_0^{\dagger 2}\left(e^{i2\phi}+1\right)+2\sqrt{\eta^2(1-\ell_\text{NL})^2}\hat{a}_0^{\dagger ' 2} e^{i\phi}\right]\ket{0}\\
    =\frac{1}{4}\left(\sqrt{\eta^2}\left(e^{i2\phi}+1\right)\ket{20}+2\sqrt{\eta^2(1-\ell_\text{NL})^2}e^{i\phi}\ket{2'0}\right),
  \end{split}
\end{equation}
with associated probability given by
\begin{equation}
  \begin{split}
    \text{P}_{20}=&\frac{1}{16}\left[\eta^2\left(2+e^{i2\phi}+e^{-i2\phi}\right)\braket{20}+4\eta^2(1-\ell_\text{NL})^2\braket{2'0}\right.\\
    &\left. +2\sqrt{\eta^4(1-\ell_\text{NL})^2}\left(\left(e^{i\phi}+e^{-i\phi}\right)\braket{2'0}{20}+\left(e^{-i\phi}+e^{i\phi}\right)\braket{20}{2'0}\right)\right]\\
    =&\frac{1}{16}\left[2\eta^2\left(1+\cos(2\phi)\right)+4\eta^2(1-\ell_\text{NL})^2+4\sqrt{\eta^4(1-\ell_\text{NL})^2}\cos(\phi)\left(r_\text{int}e^{i\theta_\text{int}}+r_\text{int}e^{-i\theta_\text{int}}\right)\right]\\ \label{eq:statistics_full}
    =&\frac{1}{16}\left[2\eta^2\left(1+\cos(2\phi)\right)+4\eta^2(1-\ell_\text{NL})^2+8\sqrt{\eta^4(1-\ell_\text{NL})^2}\cos(\phi)r_\text{int}\cos(\theta_\text{int})\right]
  \end{split}
\end{equation}

In the equations above, we have introduced the parametrization of the overlap between the input and output states from the scattering process, i.e. the overlap integral between the one- and two-photon states (Eq.~\ref{eq:input-output}), using a polar decomposition as:
\begin{equation}
  \begin{split}
    \braket{20}{2'0}=\braket{2'0}{20}^*\\
    \braket{20}{2'0}=r_\text{int}e^{i\theta_\text{int}}, \quad r_\text{int}\leq1, \quad \theta_\text{int}\in[0,2\pi]
  \end{split}
\end{equation}
which is a general description of any overlap integral between normalized quantum states. If the states are identical $\ket{20}=\ket{2'0}$, the overlap will simply be unity (losses are accounted for as prefactors of the quantum states).

\subsubsection*{Simplified parameterization of the scattering process}

We find that, for our device, a simplified parameterization of the nonlinear process can still capture the measurable photon statistics while providing a more direct interpretation. 
We define such parametrization by assuming a Kerr-type transformation for the nonlinear operations as:

\begin{equation}
   \hat{a}_0^\dagger \hat{a}_1^\dagger\rightarrow \sqrt{\eta^2(1-\ell_\text{NL})^2}\hat{a}_0^\dagger \hat{a}_1^\dagger,
    \quad \hat{a}_0^{\dagger 2} \rightarrow e^{i\varphi_\text{NL}}\sqrt{\eta^2}\hat{a}_0^{\dagger 2},
     \quad \hat{a}_1^{\dagger 2} \rightarrow e^{i\varphi_\text{NL}}\sqrt{\eta^2}\hat{a}_1^{\dagger 2}, 
\label{eq:NL_simplified}
\end{equation}
which in terms of Fock states correspond to the transformations reported in the main text:
\begin{align}
    \ket{1} &\mapsto \sqrt{\eta(1-\ell_\text{NL})} \text{e}^{i \phi} \ket{1}
    \\
    \ket{2} &\mapsto \sqrt{\eta^2} \text{e}^{i (2\phi +\varphi_\text{NL})} \ket{2}.
\end{align}

In this model, both the nonlinear phase $\theta_\text{int}$ and the finite state overlap $r_\text{int}$  are now embedded in a single parameter $\varphi_{\text{NL}}$.
The relationship between these parameters can be derived by describing the photon statistics obtainable with this simplified model and comparing it with the analogous for the full model (as in Eq.~\ref{eq:statistics_full}).
To do so, we write the evolved state according to the simplified model as
\begin{equation}
\begin{split}
    \frac{1}{2\sqrt{2}}&\left(\hat{a}_0^{\dagger2}e^{i2\phi}+\hat{a}_1^{\dagger2}+2\hat{a}_0^\dagger \hat{a}_1^\dagger  e^{i\phi}\right)\ket{0}\\
    \xrightarrow{NL}\frac{1}{2\sqrt{2}}&\left(\sqrt{\eta^2}\hat{a}_0^{\dagger2}e^{i(2\phi+\varphi_\text{NL})}+\sqrt{\eta^2}\hat{a}_1^{\dagger2}e^{i\varphi_\text{NL}}+2\sqrt{\eta^2(1-\ell_\text{NL})^2}\hat{a}_0^{\dagger} \hat{a}_1^{\dagger} e^{i\phi}\right)\ket{0}\\
    \xrightarrow{BS}\frac{1}{4\sqrt{2}}&\left[\sqrt{\eta^2}\left(\hat{a}_0^\dagger+\hat{a}_1^\dagger\right)^2e^{i(2\phi+\varphi_\text{NL})}+\sqrt{\eta^2}\left(\hat{a}_0^\dagger-\hat{a}_1^\dagger\right)^2e^{i\varphi_\text{NL}}+
    2\sqrt{\eta^2(1-\ell_\text{NL})^2}\left(\hat{a}_0^{\dagger }+\hat{a}_1^{\dagger}\right)\left(\hat{a}_0^{\dagger}-\hat{a}_1^{\dagger }\right)e^{i\phi}\right]\ket{0}\\
    =\frac{1}{4\sqrt{2}}&\left[\sqrt{\eta^2}\left(\hat{a}_0^{\dagger 2}+\hat{a}_1^{\dagger 2}+2\hat{a}_0^\dagger \hat{a}_1^\dagger\right)e^{i(2\phi+\varphi_\text{NL})}
    +\sqrt{\eta^2}\left(\hat{a}_0^{\dagger 2}+\hat{a}_1^{\dagger 2}-2\hat{a}_0^\dagger \hat{a}_1^\dagger\right)e^{i\varphi_\text{NL}}+2\sqrt{\eta^2(1-\ell_\text{NL})^2}\left(\hat{a}_0^{\dagger 2}-\hat{a}_1^{\dagger 2}\right)e^{i\phi}\right]\ket{0}.
  \end{split}
  \end{equation}
The output statistics can again be calculated easily from here, for which we obtain: 
\begin{equation}
  \begin{split}
    \text{P}_{20}=&\frac{1}{16}\left[\eta^2\left(2+e^{i2\phi}+e^{-i2\phi}\right)\braket{20}+4\eta^2(1-\ell_\text{NL})^2\braket{20}\right.\\
    &\left. +2\sqrt{\eta^4(1-\ell_\text{NL})^2}\left(\left(e^{i\phi}+e^{-i\phi}\right)e^{i\varphi_\text{NL}}\braket{20}{20}+\left(e^{-i\phi}+e^{i\phi}\right)e^{-i\varphi_\text{NL}}\braket{20}{20}\right)\right]\\
    =&\frac{1}{16}\left[2\eta^2\left(1+\cos(2\phi)\right)+4\eta^2(1-\ell_\text{NL})^2+8\sqrt{\eta^4(1-\ell_\text{NL})^2}\cos(\phi)\cos(\varphi_\text{NL})\right]
    \label{eq:statistics_simple_model}
  \end{split}
\end{equation}

\begin{figure}
    \centering
    \includegraphics{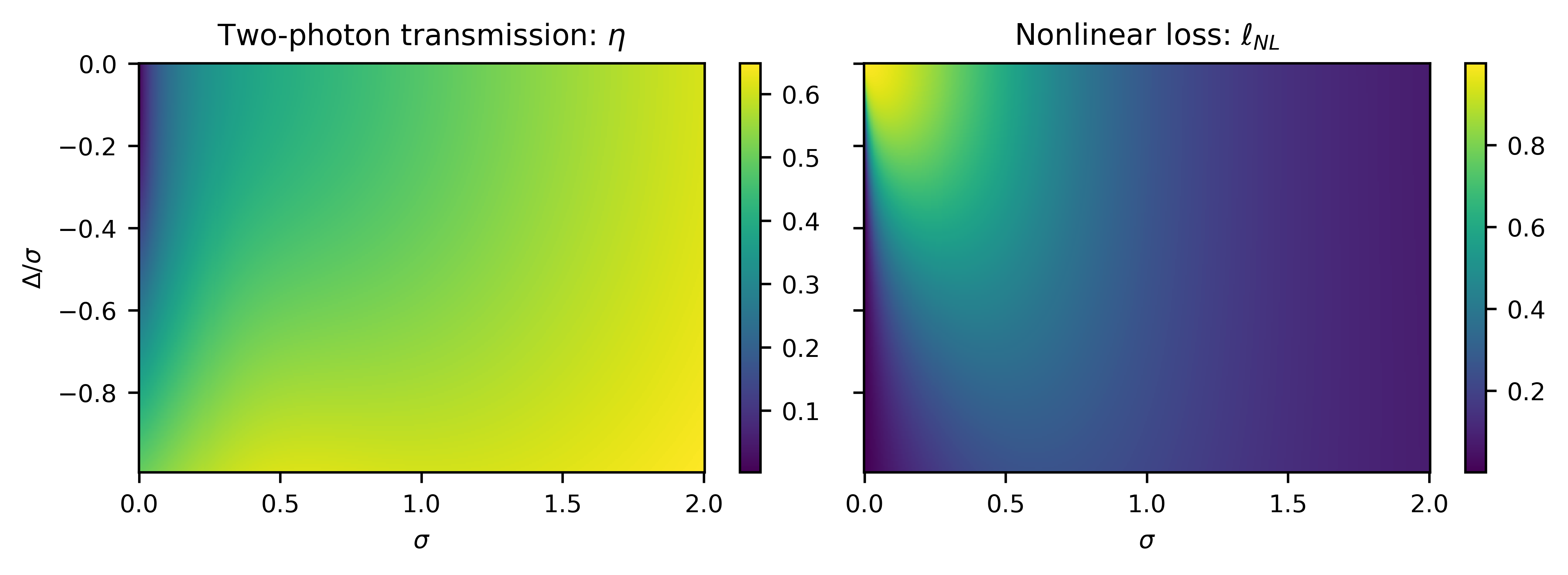}
    \caption{Theoretical values of the two-photon transmission $\eta$ and the nonlinear scattering probability $\ell_\text{NL}$ as a function of the QD lifetime $\sigma$ and relative detuning $\Delta/\sigma$.}
    \label{fig:nloss}
\end{figure}

\begin{figure}
    \centering
    \includegraphics{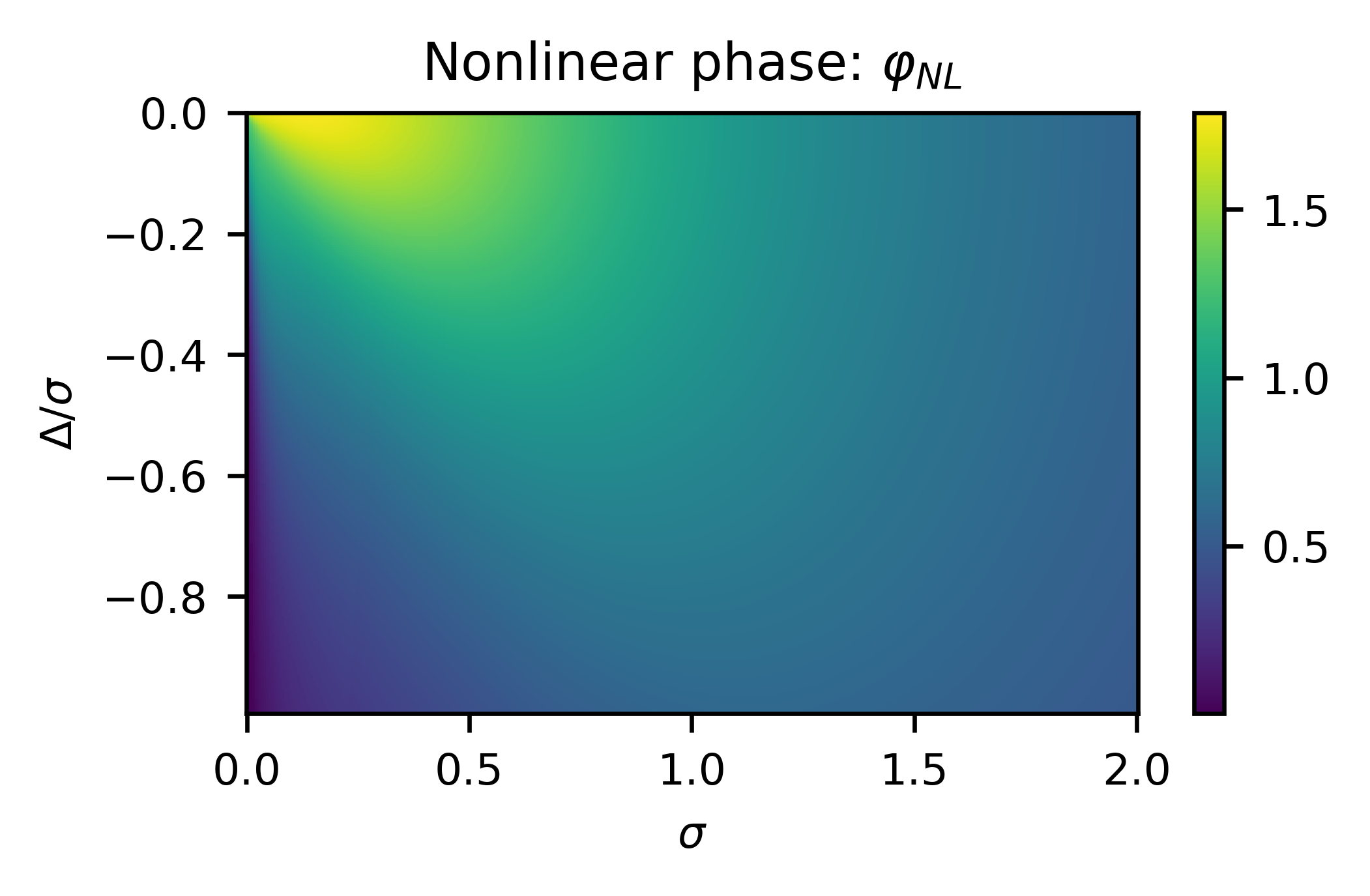}
    \caption{Theoretical values of the nonlinear phase $\varphi_\text{NL}$ as a function of the QD lifetime $\sigma$ and relative detuning $\Delta/\sigma$.}
    \label{fig:nphase}
\end{figure}

The equations describing the output statistics according to the full (Eq.~\ref{eq:statistics_full}) and the simplified (Eq.~\ref{eq:statistics_simple_model}) are equivalent via the relationship

\begin{equation}
r_\text{int}\cos(\theta_\text{int})=\cos(\varphi_\text{NL})\implies \varphi_\text{NL}=\arccos(r_\text{int}\cos(\theta_\text{int}))
    \label{eq:wf_nl_comparison}
\end{equation}
This indicates that we can transform the full wavefunction description into a simple nonlinear phase picture. Furthermore, we observe that the effective nonlinear phase applied undergoes modulation by $r_\text{int}$, which is less than one.
Note also that the sign of the interference phase $\theta_\text{int}$ alters depending on whether the detuning is positive or negative. 
However, this information is impossible to extract from the photon statistics model encompassing the nonlinear phase $\varphi_\text{NL}$ and nonlinear scattering probability $\ell_{NL}$. This limitation arises due to the symmetry nature of the setup, particularly with nonlinearities present in both modes. 
Theoretical simulations of the nonlinear parameters in our experiments according to this model are shown in Fig.~\ref{fig:nloss} and Fig \ref{fig:nphase} for the case where a perfectly coupled QD ($\beta=1$) is assumed.

\section{Details of data analysis and photon counts}

\label{appendix:analysis}

\begin{figure}
    \centering
    \includegraphics{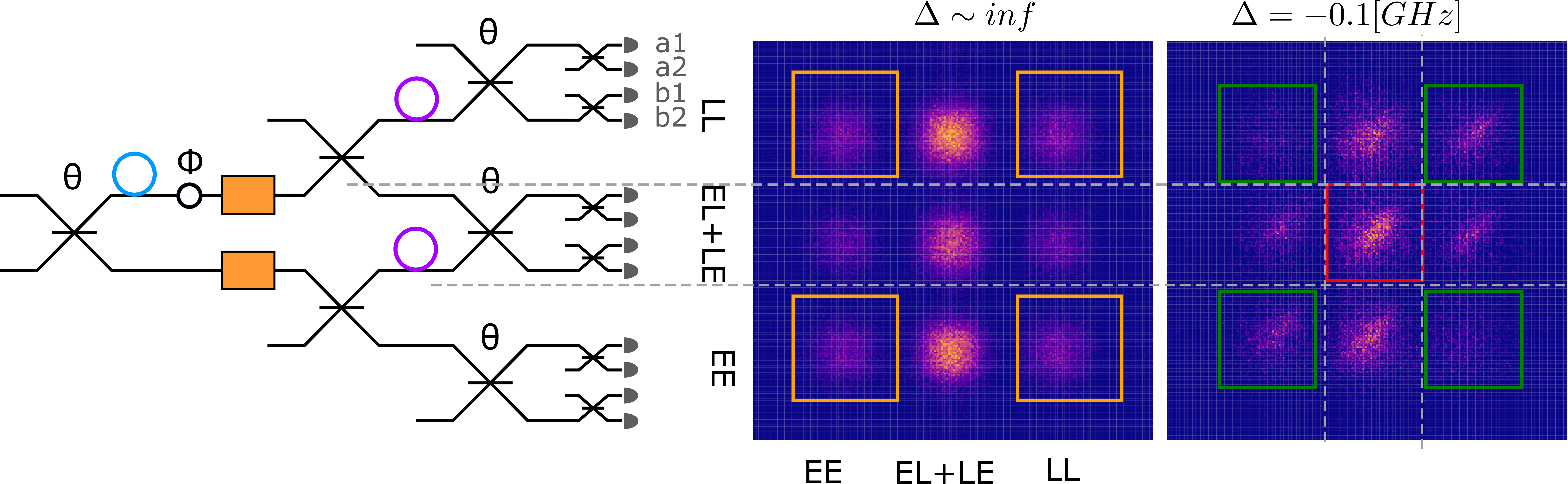}
    \caption{A schematic of the full time-bin interferometer and examples of joint time-resolved coincidence histogram. The central row (2-modes) is the only place where interference occurs. The general photon statistics for all output modes is also changed by interaction with the QD due to the unbalanced transmission coefficients.}
    \label{fig:s2}
\end{figure}

As described in Appendix~\ref{appendix:expedetails}, each of the detected output modes has 3 peaks, a central peak $m$ which is where the interference happens (due to temporal/spatial overlap) and two side peaks. When measuring the time-resolved photon coincidences, a 3x3 grid of peaks thus arises in the histogram, as seen in Fig.~\ref{fig:s2}. Assuming a two-photon input state (the lower order states are automatically filtered, and we assume no higher order states), from these histograms we can gain information about the photon scattering and interferometer alignment/coupling.
In particular, from these coincidences it is possible to probe the difference in transmission after scattering with the QD. 
In this section we describe how we do it.
A coincidence between early-early (EE) and EE or late-late(LL) and LL components can only occur if both photons go through the short path or long path twice, respectively. 
This implies that during the scattering a two-photon pulse was present. 
On the contrary, a coincidence detection in the EE and LL (or LL and EE) peaks implies that a one-photon pulse was present in both time-bin modes during the scattering. The imbalance between the number of counts in these bins thus provides the imbalance in the transmission of (squared) one-photon states and two-photon states after scattering with the QD.

The different bins also provides a characterization of unbalanced losses in the interferometer. For this, we consider coincidences between any pair of detectors to estimate the number of counts in the EE-LL (LL-EE) bins. Assuming we are far off resonance with the QD, the coincidences between $a1$-EE and $a2$-LL are be proportional to
\begin{equation}
    C_{20,\text{side}} = \frac{1}{2}\cdot\frac{1}{2}\cdot\frac{1}{2}\cdot\frac{1}{4}\cdot\frac{1}{4}\cdot\eta_E^2\eta_L^2\eta_{a1}\eta_{a2},
\end{equation}
where the factors respectively come from: one photon needs to go in each mode at the first BS  ($1/2$), at the second BS the photon that went early (late) needs to go early (late) again ($1/2^2$), and finally the photons need to go to the right combination of detectors ($1/4^2$). Finally there is the coupling of the different paths through the interferometer and of the detectors. We call the combined experiment efficiency terms for this detectors configuration $N_{20}=\eta_E^2\eta_L^2\eta_{a1}\eta_{b1}$. For the other combinations the expected number of counts in the side peaks are given by

\begin{equation}
\begin{split}
    C_{11,\text{side}} = \frac{1}{2}\cdot\frac{1}{2}\cdot\frac{1}{2}\cdot\frac{1}{4}\cdot\frac{1}{4}\cdot\eta_E^2\eta_L^2\left(\eta_{a1}\eta_{b1}+\eta_{a1}\eta_{b2}+\eta_{a2}\eta_{b1}+\eta_{a2}\eta_{b2}\right),\\
    C_{02,\text{side}} = \frac{1}{2}\cdot\frac{1}{2}\cdot\frac{1}{2}\cdot\frac{1}{4}\cdot\frac{1}{4}\cdot\eta_E^2\eta_L^2\eta_{b1}\eta_{b2},
\end{split}
\end{equation}
For the central peak (EL+LE and EL+LE), the number of counts for the (20) bunching case is given by

\begin{equation}
    C_{20,\text{center}} = \frac{1}{2}\cdot P_{(20)}\cdot\eta_E^2\eta_L^2\eta_{a1}\eta_{a2},
\end{equation}
where the factor of $(1/2)$ is due to the photons having to split up and go to different detectors in the pseudo-number-resolving scheme, $P_{(20)}$ is the probabillity for the photons to give this specific coincidence, and the last part is the efficiency, which is identical to the efficiency term in $C_{20,\text{side}}$. Note that $P_{(20)}$ is the number we are interested in calculating which depends on both the linear phase and the interaction with the QD. However, the different coincidences are globally normalized with respect to each other, meaning that the probability we are interested in calculating is
\begin{equation}
\begin{split}
    \mathcal{P}_{(20)}= & \frac{P_{(20)}}{P_{(20)}+P_{(11)}+P_{(02)}} = \frac{2C_{20,\text{center}}}{N_{20}\left(\frac{2C_{20,\text{center}}}{N_{20}}+\frac{4C_{11,\text{center}}}{N_{11}}+\frac{2C_{02,\text{center}}}{N_{02}}\right)}\\
    = & \frac{1}{1+2\frac{C_{11,\text{center}}}{C_{20,\text{center}}}\frac{N_{20}}{N_{11}}+\frac{C_{02,\text{center}}}{C_{20,\text{center}}}\frac{N_{20}}{N_{02}}} = \frac{1}{1+2\frac{C_{11,\text{center}}}{C_{20,\text{center}}}\frac{C_{20,\text{side}}}{C_{11,\text{side}}}+\frac{C_{02,\text{center}}}{C_{20,\text{center}}}\frac{C_{20,\text{side}}}{C_{02,\text{side}}}},
\end{split}
\end{equation}
where the efficiency factors can be directly exchanged with the counts in the side peaks due to the prefactors being the same for all three cases. The factor of 2 on the (11) antibunching term is due to a factor of $1/4$ rather than $1/2$ in front of the $C_{11,\text{center}}$ term.

\section{Tests for the model for photon statistics}
\label{appendix:photonstatistics}

\begin{figure}      
    \includegraphics[width=.48\linewidth]{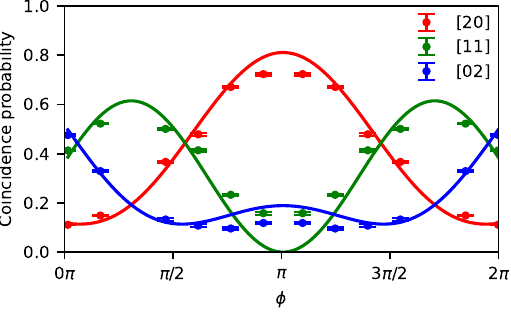}
    \includegraphics[width=.48\linewidth]{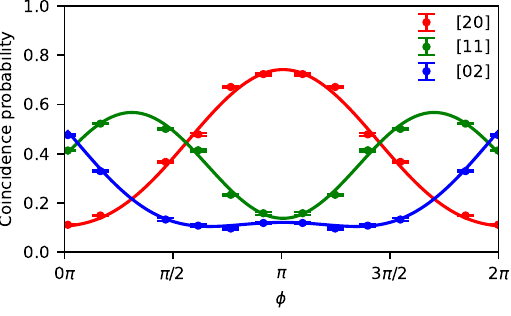}
    \caption{Models for photon statistics. Left: Quantum interference data fitted to the model for two-photon statistics including the linear and nonlinear contributions, but not distinguishability effects. Right: same but including photon distinguishability in the model for the fits.}
    \label{fig:S_ideal_stat_model}
\end{figure}

In Appendix~\ref{appendix:input_output} the photon coincidence statistics are given for the two-photon input into a MZI with a QD situated in both arms. It is also clear that the general loss $\eta$ is not relevant for the overall statistics since it can be factored out. Thus the only relevant parameters are the linear phase, the nonlinear phase and the nonlinear loss
\begin{equation}
  \begin{split}
    \text{P}_{20}=& \frac{1}{16}\eta^2\left[2\left(1+\cos(2\phi)\right)+4(1-\ell_\text{NL})^2+8(1-\ell_\text{NL})\cos(\phi)\cos(\varphi_\text{NL})\right],\\
    \text{P}_{02}=& \frac{1}{16}\eta^2\left[2\left(1+\cos(2\phi)\right)+4(1-\ell_\text{NL})^2-8(1-\ell_\text{NL})\cos(\phi)\cos(\varphi_\text{NL})\right],\\
    \text{P}_{11}=& \frac{1}{4}\eta^2\left[1-\cos(2\phi)\right],
\end{split}
\end{equation}
since the global probabilities are renormalized after detection. An example of fitting this model to the data is shown in Fig.~\ref{fig:S_ideal_stat_model}.
The fits describe the observed statistics in most cases, with spurious variations (e.g. the probability for (11) coincidence not going to zero) which can be accounted to additional noise processes not included in the model, such as partial distinguishability between the photons from the two arms of the interferometer. Including the partial distinguishability in the fit model shown in Fig.~\ref{fig:S_ideal_stat_model} right, gives a better fit compared to the model without. The partial distinguishability of 10 \% is obtained in the model which is comparable to the value measured from the resonant fluorescence \ref{fig:S_life_g2_HOM}.
With the QD off the (11) probability goes to zero, thus the effect does not arise from the linear optics, but rather from the scattering process itself. This indicates that the two scatterings are not identical.

Spectral diffusion could be an explanation for this e.g. the resonance frequency of the QD drifts between the two scatterings. However the delay between the early and late scattering is on the order of 5ns, which is too fast for any substantial spectral diffusion to take place. Instead this effect is probably attributed to some pure dephasing process, which acts on these time-scale. 

In order to model this effect, we introduce orthogonal ancillary virtual modes in the description, and make a unitary coupling to this extra mode. This model can describe partial distinguishability between the early and late scattering~\cite{Tichy2015}
\begin{equation}
    \hat{a}_0^\dagger \xrightarrow{\perp} \cos{(\theta_\perp)} \hat{a}_0^\dagger+ \sin{(\theta_\perp)}\hat{a}_{0\perp}^\dagger,
\end{equation}
where the $\hat{a}_1^\dagger$ would not go through the same transformation but stay in the initial mode. Implementing this additional unitary transformation to the equations for the two-photon statistics from section~\ref{appendix:from_input_output_to_nl} we arrive at the following
\begin{align}
\begin{split}
    \frac{1}{2\sqrt{2}}&\left(\sqrt{\eta^2}\hat{a}_0^{\dagger2}e^{i(2\phi+\varphi_\text{NL})}+\sqrt{\eta^2}\hat{a}_1^{\dagger2}e^{i\varphi_\text{NL}}+2\sqrt{\eta^2(1-\ell_\text{NL})^2}\hat{a}_0^{\dagger} \hat{a}_1^{\dagger} e^{i\phi}\right)\ket{0},\\
    \xrightarrow{\perp}
    \frac{1}{2\sqrt{2}}&\left[\sqrt{\eta^2}\left(\cos{(\theta_\perp)}\hat{a}_0^{\dagger}+\sin{(\theta_\perp)}\hat{a}_{0\perp}^{\dagger}\right)^2e^{i(2\phi+\varphi_\text{NL})}+\sqrt{\eta^2}\hat{a}_1^{\dagger2}e^{i\varphi_\text{NL}}\right.\\
    &\left.+2\sqrt{\eta^2(1-\ell_\text{NL})^2}\left(\cos{(\theta_\perp)}\hat{a}_0^{\dagger}+\sin{(\theta_\perp)}\hat{a}_{0\perp}^{\dagger}\right) \hat{a}_1^{\dagger} e^{i\phi}\right]\ket{0},\\
    =\frac{1}{2\sqrt{2}}&\left[\sqrt{\eta^2}\left(\cos^2{(\theta_\perp)}\hat{a}_0^{\dagger2}+\sin^2{(\theta_\perp)}\hat{a}_{0\perp}^{\dagger2}+2\cos{(\theta_\perp)}\sin{(\theta_\perp)\hat{a}_0^{\dagger}\hat{a}_{0\perp}^{\dagger}}\right)e^{i(2\phi+\varphi_\text{NL})}\right.\\
    &\left.+\sqrt{\eta^2}\hat{a}_1^{\dagger2} e^{i\varphi_\text{NL}}+2\sqrt{\eta^2(1-\ell_\text{NL})^2}\left(\cos{(\theta_\perp)}\hat{a}_0^{\dagger}+\sin{(\theta_\perp)}\hat{a}_{0\perp}^{\dagger}\right) \hat{a}_1^{\dagger} e^{i\phi}\right]\ket{0},\\
    \xrightarrow{BS}\frac{1}{4\sqrt{2}}&\left[\sqrt{\eta^2}\left(\cos^2{(\theta_\perp)}(\hat{a}_0^{\dagger}+\hat{a}_1^{\dagger})^2+ \sin^2{(\theta_\perp)}(\hat{a}_{0\perp}^{\dagger}+\hat{a}_{1\perp}^{\dagger})^2 +2\cos{(\theta_\perp)}\sin{(\theta_\perp)}(\hat{a}_0^{\dagger}+\hat{a}_0^{\dagger})(\hat{a}_{0\perp}^{\dagger}+\hat{a}_{1\perp}^{\dagger})\right)e^{i(2\phi+\varphi_\text{NL})}\right.\\
    &\left.+\sqrt{\eta^2}(\hat{a}_0^{\dagger}-\hat{a}_1^{\dagger})^2e^{i\varphi_\text{NL}}+2\sqrt{\eta^2(1-\ell_\text{NL})^2}\left(\cos{(\theta_\perp)}(\hat{a}_0^{\dagger}+\hat{a}_1^{\dagger})+\sin{(\theta_\perp)}(\hat{a}_{0\perp}^{\dagger}+\hat{a}_{1\perp}^{\dagger})\right) (\hat{a}_0^{\dagger} -\hat{a}_1^{\dagger} )e^{i\phi}\right]\ket{0},
\end{split}
\end{align}
from which the two-photon statistics can be calculated. Note that $\bra{0}\hat{a}_{0\perp}\hat{a}_{0}^{\dagger}\ket{0}=0$.

\section{Quantum simulation of water molecule}
\label{appendix:water}
\subsection{Anharmonic water}

\begin{figure}
    \centering
    \includegraphics{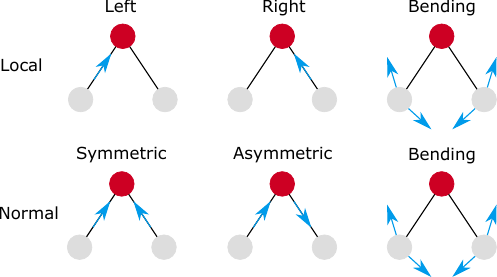}
    \caption{A schematic of the three vibrational modes in a water molecule. Upper part shows the local basis and the lower part shows the normal or eigenbasis.}
    \label{fig:vibrational_modes}
\end{figure}

The $\text{H}_2 \text{O}$ water molecule has three phononic modes, which are described as coupled anharmonic oscillators. In the local basis these are the \textit{left}, \textit{right}, and \textit{bending} modes, shown in the top panel of Fig.~\ref{fig:vibrational_modes}. However, the coupling between the bending mode and the stretch modes is negligible. In the experiment, the simulation procedure starts by transforming from the local basis to the normal(eigen) modes of the Hamiltonian, given by the symmetric and antisymmetric stretch modes shown in the bottom panel of Fig.~\ref{fig:vibrational_modes}. 

The frequencies of the single excitations of the eigenmodes are given by~\cite{mizukami2013,sparrow2018}
\begin{equation}
    \nu_{\ket{10}=\text{Sym.}} = 3740.05 \text{cm}^{-1}, \quad \nu_{\ket{01}=\text{Asym.}} = 3619.68 \text{cm}^{-1},
\end{equation}
whereas the frequencies for two excitations are given 
\begin{equation}
  \nu_{\ket{20}} = 7391.43 \text{cm}^{-1}, \quad \nu_{\ket{02}} = 7154.35 \text{cm}^{-1}, \quad \nu_{\ket{11}} = 7206.46 \text{cm}^{-1},
\end{equation}
from which the anharmonicity is clearly seen, i.e. $\nu_{\ket{20}} \neq 2\nu_{\ket{10}}$.

For the $\text{H}_2 \text{O}$ molecule, this basis transformation from the local stretch modes to the eigenmodes corresponds to a unitary transformation equivalent to that of a 50:50 beam splitter.
\begin{equation}
  U_L=\frac{1}{\sqrt{2}}\mqty(1 & -1 \\ 1 & 1).
\end{equation}

\end{document}